\documentclass[twocolumn,showkeys,showpacs,pre]{revtex4}
\usepackage{amsfonts, amsmath, latexsym}
\usepackage[dvips]{graphicx}

\newtheorem{1}{Theorem}
\newtheorem{2}[1]{Theorem}
\newtheorem{3}[1]{Theorem}
\newtheorem{4}[1]{Theorem}
\newtheorem{5}[1]{Theorem}
\newtheorem{6}{Corollary}

\begin{document}

\title{Random Series and Discrete Path Integral methods: The L\'evy-Ciesielski implementation}

\author{Cristian Predescu} 
\email{Cristian_Predescu@brown.edu}
\author{J. D. Doll}
 
\affiliation{
Department of Chemistry, Brown University, Providence, Rhode Island 02912
}
\date{\today}
\begin{abstract}
We perform a thorough analysis of the relationship between discrete and series representation path integral methods, which are the main numerical techniques used in connection with the Feynman-Ka\c{c} formula.  First, a new interpretation of the so-called standard discrete path integral methods is derived by direct discretization of the Feynman-Ka\c{c} formula. Second, we consider a particular random series technique based upon the L\'evy-Ciesielski representation of the Brownian bridge and  analyze its main implementations, namely the primitive, the partial averaging, and the reweighted versions. It is shown that the $n=2^k-1$ subsequence of each of these methods can also be interpreted as a discrete path integral method with appropriate short-time approximations. We therefore establish a direct connection between the discrete and the random series approaches. In the end, we give sharp estimates on the rates of convergence of the partial averaging and the reweighted L\'evy-Ciesielski random series approach for sufficiently smooth potentials. The asymptotic rates of convergence are found to be $\mathcal{O}(1/n^2)$, in agreement with the  rates of convergence of the best standard discrete path integral techniques. 
\end{abstract}
\maketitle

\section{Introduction} 
\newcommand{\ud}{\mathrm{d}}
Ever since their introduction over fifty years ago, path integral methods have been an intense research field for physicists, chemists, and mathematicians alike, even if these researchers have sometimes used arguments of a rather different nature. The field began with Feynman's observation \cite{Fey48} that the time propagator of the Schr{\"o}dinger equation can be represented as a ``sum over histories,'' effectively giving a formula for the propagator as a limit of integrals over spaces of increasing dimension \cite{Fey65}. In mathematical terms, the existence of this limit is problematic though several research directions are known \cite{Alb76, Joh00, Car95, deF91}.

 In a significant development, Ka\c{c} noticed that the ``imaginary time'' version of the formula utilized by Feynman has a definite probabilistic sense and  could be interpreted as an integral of a functional of the seemingly ubiquitous Brownian motion \cite{Kac51}. Such a formula could represent the Green's function for a certain class of diffusion processes, as for instance the density matrix for the Bloch equation. The end product of their work is beautifully summarized by what is now called the  Feynman-Ka\c{c} formula \cite{Sim79}  
\begin{equation}
\label{eq:1}
\frac{\rho(x,x';\beta)}{\rho_{fp}(x,x';\beta)}=\mathbb{E}\exp\left\{-\beta\int_{0}^{1}\! \!  V\Big[x_r(u)+\sigma B_u^0 \Big]\ud u\right\},
\end{equation}
where $\rho(x,x';\beta)$ is the density matrix for a monodimensional canonical system characterized by the inverse temperature $\beta=1/(k_B T)$ and made up of identical particles of mass $m_0$  moving in the potential $V(x)$. 
The stochastic element that appears in Eq.~(\ref{eq:1}), $\{B_u^0,\, u\geq
0\}$, is a so-called standard Brownian bridge defined as follows: if  $\{B_u,\, u\geq
0\}$ is a standard Brownian motion starting at zero, then the Brownian
bridge is the stochastic process~$\{B_u |\,B_1=0,\, 0 \leq u \leq 1\}$
i.e., a Brownian motion conditioned on~$B_1=0$ \cite{Dur96}. In this
paper, we shall reserve the symbol~$\mathbb{E}$ to denote the expected
value (average value) of  a certain random variable against the
underlying probability measure of the Brownian bridge~$B_u^0$. To complete the description of Eq~(\ref{eq:1}), we set $x_r(u)=x+(x'-x)u$ (called the reference path), $\sigma= (\hbar^2\beta  /m_0)^{1/2}$, and let $\rho_{fp}(x,x';\beta)$ denote the density matrix for a similar free particle. 

Rather than directly employing Eq.~(\ref{eq:1}), chemical physicist's arguments are usually constructed around the Trotter composition rule \cite{Tro59} that exploits the fact that $\{e^{-\beta H}; \beta > 0\}$ is a semigroup of operators on $L^2(\mathbb{R})$, so that 
\begin{equation}
\label{eq:2}
e^{-(\beta_1 +\beta_2)H}= e^{-\beta_1 H}e^{-\beta_2 H}
\end{equation} 
or, in coordinate representation,
\begin{equation}
\label{eq:3}
\langle x|e^{-(\beta_1 +\beta_2)H}| x' \rangle= \int_{\mathbb{R}} \ud z \langle x| e^{-\beta_1 H}|z\rangle \langle z|e^{-\beta_2 H}|x'\rangle.
\end{equation} 
By  writing $\beta= \sum_{k=1}^n \beta_k$, repeatedly applying the Trotter rule and choosing an adequate ``short-time'' approximation, one ends up with a sequence of integrals on spaces of increasing dimension, converging to the density matrix as $\max_{1\leq k\leq n} \beta_k \to 0$. Of course, this is much in the spirit of the original Feynman path integral approach. The methods deduced by this technique are usually called Discrete Path Integral (DPI) methods (see Ref.~\onlinecite{Mie01} and the cited bibliography).

It has become apparent that the Ka\c{c} interpretation of the Feynman's formula may, in fact, offer a valuable starting point for the general construction finite dimensional approximations to the density matrix. This is so because the Brownian motion is a mathematically well understood object, for which various constructions are known. For example, the use of the Ito-Nisio theorem \cite{Kwa92} has lead the present authors to the development of the random series path integral methods in a surprisingly general fashion \cite{Pre02}. This generality of the theory allows one to identify optimality criteria and eventually to answer questions as what the best representation is or how to modify the approach in order to improve the convergence. 

In this paper, we shall look at the relation between the Ka\c{c} interpretation of the Feynman formula and discrete path integral methods. In the first part, we consider what we call the standard DPI methods. We shall again show  the strength of the Ka\c{c} approach, at least in terms of generality and mathematical interpretation of the formulae. In the second part, we explore the connection between the random series technique, as particularized for the L\'evy-Ciesielski series representation of the Brownian bridge, and certain DPI results from the chemical literature.   While not the primary goal, we do obtain in an effortless manner a  better way of implementing the latter results that features a ``built in'' fast sine-Fourier transform. We shall also derive the two basic modifications of the L\'evy-Ciesielski path integral method (LCPI), the partial averaging \cite{Dol85,Pre02} and the reweighted techniques \cite{Pre02}, and establish their asymptotic law of convergence. We suggest that these results again emphasize the power of the Ka\c{c} interpretation of the Feynman formula.  By providing a central framework for discussion and analysis, the Ka\c{c} approach significantly aids in characterizing the various methods and in establishing their interconnections, links that otherwise would be obscured by the multitude of possible representations.

\section{The standard Discrete Path Integral method}
\subsection{Trotter-Suzuki approach.}
Our definition of the standard DPI method has to do with the particular short-time approximation that replaces the exact one in the Trotter product
\begin{equation}
\label{eq:4a}
e^{-\beta H}=\underbrace{e^{-\frac{\beta}{n+1}H}\ldots e^{-\frac{\beta}{n+1}H}}_{n+1 \,\text{terms}}.
\end{equation}
We follow closely the arguments of Suzuki \cite{Suz91}. The Hamiltonian of the system can be written as a sum between  the kinetic operator and the potential operator in the form $H=K+V$. The coordinate representations for the two operators are analytically known to be
\[\langle x| e^{-\beta K}|x'\rangle = \rho_{fp}(x,x';\beta)\]
and
\[\langle x| e^{-\beta V}|x'\rangle= e^{-\beta V(x)}\delta(x'-x),\]
respectively. It is therefore natural to consider short-time approximations that can be expressed by a finite composition of the above density matrices. The simplest example is the $2$-term splitting formula
\begin{equation}
\label{eq:4}
e^{-\beta(K+V)}=e^{-\beta K}e^{-\beta V}[1+\mathcal{O}(\beta^2)],
\end{equation}
which is of order $1$. More generally, the order of a splitting formula is said to be $k$ if the relative error is $\mathcal{O}(\beta^{k+1})$.  The motivation for this is  that if 
\[e^{-\beta(K+V)}=f_k(K,V;\beta)[1+\mathcal{O}(\beta^{k+1})],\]then 
\begin{equation}
\label{eq:5}
e^{-\beta(K+V)}=\left[f_k\left(K,V;\frac{\beta}{n}\right)\right]^n \left[1+\mathcal{O}\left(\frac{\beta^{k+1}}{n^k}\right)\right]
\end{equation}
i.e., the error of the final $n$-term Trotter product formula decays as fast as $1/n^k$. The relation (\ref{eq:5}) was actually proved by Suzuki \cite{Suz85} in terms of operator norms for bounded operators $A$ and $B$, but such an estimation also holds for $K$ and $V$ (which generally are unbounded operators; however, $e^{-\beta(K+V)}$ is bounded for most of the potentials of physical interest and for all positive $\beta$).

A better splitting is offered by the $3$-term formula 
\begin{equation}
\label{eq:6}
e^{-\beta(K+V)}=e^{-\frac{1}{2}\beta V}e^{-\beta K}e^{-\frac{1}{2}\beta V}[1+\mathcal{O}(\beta^3)],
\end{equation}
or the one obtained by permuting $V$ with $K$. These are of order $2$ and go by the name of symmetrical trapezoidal Trotter short-time approximations \cite{Rae83, Suz91}. More generally, let us define a $2l+1$-term splitting formula of order $k$ by the expression
\begin{eqnarray}
\label{eq:7}
e^{-\beta (K+V)}\nonumber= &&e^{-a_0 \beta V}e^{-b_1 \beta K}e^{-a_1 \beta V}\ldots \\ &&\ldots e^{-b_l \beta K} e^{-a_{l} \beta V} [1+\mathcal{O}(\beta^{k+1})]. 
\end{eqnarray} 
Symmetry arguments suggest that for the  optimum $2l+1$-term splitting formula the sequences $a_0, \ldots, a_l$ and $b_1, \ldots, b_l$ should be \emph{palindromic} i.e., if the coefficients are read left to right, they form the same numerical sequence as if they are read right to left. A look at the trapezoidal Trotter formula shows that this condition is natural, as one has little reason to believe that anything new can be achieved by considering some arbitrary 
$e^{-a\beta V}e^{-\beta K}e^{-(1-a)\beta V}$ decomposition. In fact, with the help of the Campbell-Baker-Haussdorf-Dynkin formula \cite{Var88}, it can be shown that this more general expression is an order $2$ splitting only if $a=1/2$ and that it is an order $1$ splitting otherwise. More generally, since the operator $e^{-\beta H}$ is Hermitian, it is natural and, as argued by De~Raedt and De~Raedt \cite{Rae83}, optimal  to approximate it by a sequence of Hermitian operators. It is straightforward to see that the $n$-order Trotter product (\ref{eq:4a}) is Hermitian if and only if the short-time approximation (\ref{eq:7}) is Hermitian.  In turn, this requires the palindromicity of the $\{a_i\}$ and $\{b_i\}$ sequences. 

It is not difficult to see that if $\sum_i a_i =p$ and $\sum_i b_i =q$, then the $n$-order Trotter formula (\ref{eq:4a}) converges to $\exp[-\beta( q K+p V)]$. On the other hand, the equality \[e^{-\beta( q K+p V)}=e^{-\beta( K+ V)}\] holds for arbitrary potentials $V(x)$ if and only if $q=1$ and $p=1$. Therefore, the additional constraints $ \sum_i a_i =1$ and $\sum_i b_i =1$ must be enforced upon the sequences $\{a_i\}$ and $\{b_i\}$.

We considered this more general problem with the hope that by using a more advanced splitting  one may improve the asymptotic order of the Trotter product formula. Now, there is one more  restriction that we have to place on the sequence $a_0, b_1, a_1, \ldots$, namely it should be made up of \emph{real and positive} numbers only. Otherwise, the short-time approximations are either ill-defined or, by Trotter composition, generate algorithms that are numerically unstable at low  temperatures.  Unfortunately, the following theorem of Suzuki (see Theorem~3 of Ref.~\onlinecite{Suz91}) says that 
\begin{1}[Suzuki nonexistance theorem]
There are no finite length splitting formulae (\ref{eq:7}) of order $3$ or more such that the coefficients  $a_0, b_1, a_1, \ldots$ are all real and positive.
\end{1}

This negative result shows that more general splitting formulas do not produce short-time approximations capable of improving upon the trapezoidal Trotter result, at least as far as the asymptotic order of the Trotter product rule is concerned. However, the product rule (\ref{eq:4a}), which uses equally spaced time slices, does not provide the most general standard DPI expression.  In the next section, we shall argue that this most general expression is of the form given by Eq.~(\ref{eq:7}), for which the Suzuki nonexistance theorem does not apply. 

\subsection{Direct quadrature of the Feynman-Ka\c{c} formula.}

Let us notice that the form of the equation (\ref{eq:7}) is invariant under the Trotter composition (\ref{eq:4a}) and so it can be regarded as the most general standard DPI approximation to the density matrix provided that we can give a recipe for choosing the sequences $a_0, b_1, a_1, \ldots, b_l, a_l$ in such a way that the correct result is recovered in the limit $l\to \infty$.  While in the Trotter-Suzuki approach this may seem a daunting task, the problem has an easy solution by means of the Ka\c{c} interpretation of the Feynman formula. In this section, we shall derive a  more general expression for the standard Discrete Path Integral method simply by replacing the monodimensional integral over $u$ in Eq.~(\ref{eq:1}) with an approximate quadrature sum and then using the definition of the Brownian bridge to compute the expectation of the resulting functional. Given the Suzuki nonexistence theorem, it is hard to believe that one may eventually devise a standard DPI method with asymptotic convergence $\mathcal{O}(1/n^3)$ or better. However, before one starts to investigate the validity of this conjecture, one needs a more  general  statement of the standard DPI method.   

For obvious reasons, the random process $W_{x,x'}^\sigma(u)=x_r(u)+\sigma B_u^0$ is called  a Brownian bridge of variance $\sigma^2$ and end points $(x,x')$. 
The Feynman-Ka\c{c} formula can be expressed in terms of the new process in the form 
\begin{equation}
\label{eq:8}
\frac{\rho(x,x';\beta)}{\rho_{fp}(x,x';\beta)}=\mathbb{E}\exp\left\{-\beta\int_{0}^{1}\! \!  V\Big[W_{x,x'}^\sigma(u) \Big]\ud u\right\}.
\end{equation}
A quite important property of the Brownian bridge $W_{x,x'}^\sigma(u)$  is the joint distribution of the variables $ W_{x,x'}^\sigma(u_1), \ldots, W_{x,x'}^\sigma(u_n)$ for a given partitioning $0< u_1<\ldots <u_n<1$ of the interval $[0,1]$. Let us set \[p_t(x)=\frac{1}{\sqrt{2\pi t}}e^{-{x^2}/{2t}}\] and notice that $\rho_{fp}(x,x';\beta)=p_{\sigma^2}(x'-x).$ From the very definition of the Brownian motion \cite{Dur96a}, the aforementioned joint distribution can be straightforwardly shown to be
\begin{widetext}
\begin{eqnarray}
\label{eq:9}
P\left\{ W_{x,x'}^\sigma(u_1)\in [x_1, x_1+\ud x_1], \ldots,  W_{x,x'}^\sigma(u_n)\in [x_n, x_n+\ud x_n]\right\}\nonumber \\= p_{\sigma^2 u_1}(x_1-x)p_{\sigma^2(u_2-u_{1})}(x_2-x_{1}) \ldots p_{\sigma^2(u_n-u_{n-1})}(x_n-x_{n-1}) \\ \times p_{\sigma^2(1-u_n)}(x'-x_n)\big/ p_{\sigma^2}(x'-x)\;\ud x_1 \ldots \ud x_n. \nonumber
\end{eqnarray}
\end{widetext}
One may use the above joint distribution density to compute the expectations of the functionals of the Brownian bridge which are of the form 
\begin{widetext}
\begin{eqnarray}
\label{eq:10}
p_{\sigma^2}(x'-x)\mathbb{E}\{f[W_{x,x'}^\sigma(u_1), \ldots, W_{x,x'}^\sigma(u_n)]\}=\int_{\mathbb{R}}\ud x_1\ldots \int_{\mathbb{R}}\ud x_n \; f(x_1,\ldots, x_n)\nonumber \\ \times p_{\sigma^2 u_1}(x_1-x)p_{\sigma^2 (u_2-u_{1})}(x_2-x_{1})\ldots p_{\sigma^2 (u_n- u_{n-1})}(x_n-x_{n-1}) p_{\sigma^2(1-u_n)}(x'-x_n),
\end{eqnarray}
\end{widetext}
where $f(x_1, \ldots, x_n)$ is some integrable $n$-dimensional function. 
As a direct application of Eq.~(\ref{eq:10}), consider a quadrature scheme on the interval $[0,1]$ specified by the points $0=u_0 < u_1 <\ldots <u_n < u_{n+1}=1$ and the corresponding \emph{nonnegative} weights $w_0, w_1, \ldots, w_{n+1}$. Replacing the monodimensional integral in the Feynman-Ka\c{c} formula (\ref{eq:8}) by its quadrature form, we obtain an approximation to the density matrix of the form 
\begin{equation}
\label{eq:11}
\frac{\rho_{n}^{\text{DPI}}(x,x';\beta)}{p_{\sigma}(x'-x)}=\mathbb{E}\exp\left\{-\beta\sum_{i=0}^{n+1} w_i V\Big[W_{x,x'}^\sigma(u_i) \Big]\ud u\right\}.
\end{equation}
The expectation value of this formula can be \emph{exactly} reduced to a finite dimensional integral with the help of the formula (\ref{eq:10}). We call Eq.~(\ref{eq:11}) the standard Discrete Path Integral (DPI) method and we expect it to converge to the correct result for all continuous and bounded from below potentials $V(x)$. In this respect, remember that with probability one the Brownian paths are continuous and therefore so is  $V[W_{x,x'}^\sigma(u) ]$ as a function of $u$. Also remember that by definition a quadrature scheme on $[0,1]$ is constructed so that it eventually integrates all continuous functions on $[0,1]$. 

Formula (\ref{eq:11}) can indeed be formally deduced starting with the Trotter composition rule (\ref{eq:2}) and a carefully chosen sequence of short-time approximations. More precisely, for $i=0,1\ldots, n$,  define $\theta_i=u_{i+1}-u_i$. Then the equation (\ref{eq:11}) is nothing else but the Trotter product 
\begin{eqnarray}
\label{eq:12}
&&e^{-w_0 \beta V}e^{-\theta_0 \beta K}e^{-w_1 \beta V}e^{-\theta_1 \beta K} \ldots \nonumber \\ &&\ldots e^{-w_n \beta V} e^{-\theta_{n} \beta K} e^{-w_{n+1} \beta V},  
\end{eqnarray} 
which is of course of the type given by the formula (\ref{eq:7}).  Finally, let us notice that we always have $\sum_i \theta_i=1$. We also require that $\sum_i w_i=1$.  In fact, one is not interested in relaxing these equalities because they are the necessary and sufficient conditions to obtain the exact free particle density matrix at all levels of approximation. The reader can directly verify this fact by assuming that the potential $V(x)$ in Eq.~(\ref{eq:12}) is constant but not zero. Moreover, the additional restriction of the integration schemes to those for which  the sequences $\theta_i$ and $w_i$ are palindromic is justified by the requirement that the approximate density matrices be Hermitian.

To summarize, the advantage of the equation (\ref{eq:11}) is the novel interpretation for the sequences $\theta_i=u_{i+1}-u_i$ and $w_i$, leading us to a more general convergence problem: what is the best convergence order for the standard DPI approach (\ref{eq:11}) and for what types of quadrature schemes is it attained? As we suggested in the beginning of this subsection, it is very plausible that the answer to the above question is two and is attained for almost all ``sensible'' quadrature schemes.  We illustrate this by studying the convergence of the diagonal matrix element $\rho(0;\beta)=\langle 0| e^{-\beta H}|0 \rangle$ of an harmonic oscillator for the following quadrature techniques: the trapezoidal rule (TT) and the Gauss-Legendre method (GL) \cite{Pre92}. In both cases the condition $\sum_i w_i=1$ and the palindromicity of the sequences $\theta_i$ and $w_i$ are respected.  We leave it for the reader to show that if the trapezoidal rule is used for integration, then one recovers the classical trapezoidal Trotter formula. 

If $Mt$ stands for any of the methods studied and if $\alpha_{Mt}$ represents the convergence order of the corresponding matrix element $\rho_n^{Mt}(0;\beta)$, then the convergence constant is defined by 
\[
c_{Mt}=\lim_{n \to \infty}n^{\alpha_{Mt}}[\rho(0;\beta)-\rho_n^{Mt}(0;\beta)].
\]
The above relation can be cast in the more intuitive but equivalent form \[\rho(0;\beta)\approx\rho_n^{Mt}(0;\beta)+ \frac{c_{Mt}}{n^{\alpha_{Mt}}},\]
with an appropriate definition of the  symbol $\approx$. 
These convergence orders and convergence constants can be evaluated numerically as follows. For each method, we compute 
\[
\alpha^{Mt}_n=(n^2-1/4)\log\left[1+\frac{\rho_{4n-2}^{Mt}(0;\beta)-\rho_{4n+2}^{Mt}(0;\beta)}{\rho_{4n+2}^{Mt}(0;\beta)-\rho(0;\beta)}\right],
\]
where $\rho_{4n+2}^{Mt}(0;\beta)$ represents the DPI approximation of order $4n+2$  for the method $Mt$. The evaluation of the matrix elements $\rho_{4n+2}^{Mt}(0;\beta)$  is discussed in Appendix~B. Then, as argued in Ref.~(\onlinecite{Pre02}), $\alpha^{Mt}_n$ as a function of $n$ is asymptotically a straight line, whose slope gives the convergence order. Therefore, $\alpha_{Mt}=\lim_{n \to \infty}\alpha^{Mt}_{n+1}-\alpha^{Mt}_n$. As to the convergence constants, they can be evaluated by studying the asymptotic slopes of 
\[
c^{Mt}_n=(4n+2)^{\alpha_{Mt}} (n+1/2)\left[{\rho_{4n+2}^{Mt}(0;\beta)-\rho(0;\beta)}\right],
\]
once $\alpha_{Mt}$ is known.
The computations were performed in atomic units for a particle of mass $m_0=1$ and for the harmonic oscillator $V(x)=x^2/2$. The inverse temperature was $\beta=10$. 
\begin{figure}[!tbp] 
  \includegraphics[angle=270,width=8.5cm,clip=t]{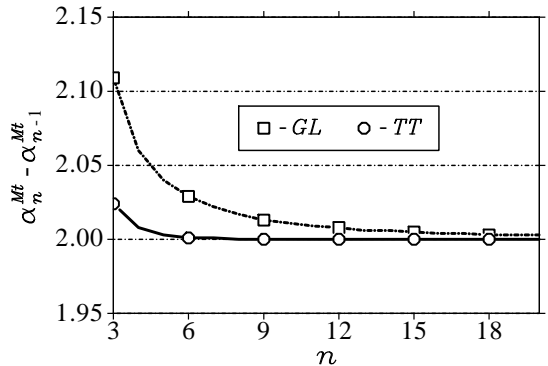} 
 \caption[sqr]
{\label{Fig:1}
The current slopes $\alpha^{Mt}_{n+1}-\alpha^{Mt}_n$ for the trapezoidal rule (TT) and for the Gauss-Legendre quadrature method (GL) are shown here to converge to the same value of $2$. 
}
\end{figure}
As shown in Fig.~\ref{Fig:1}, the asymptotic convergence order of both methods is 2. The convergence constants are found to be $c_{TT}=0.103$ and $c_{GL}=0.127$, respectively. One notices that at the temperature $\beta=10$, the trapezoidal Trotter method is slightly faster. However for $\beta=1$, one computes $c_{TT}=0.033$ and $c_{GL}=0.005$, which indicates that for this temperature the Gauss Legendre method is faster. 

The conclusion we draw from this analysis is that ``better'' integration schemes do not necessarily improve upon the convergence of the standard DPI methods. Why is this so? Again the Ka\c{c} interpretation of the Feynman formula gives us an explanation which is not obvious from the Trotter composition rule. A famous theorem due to Paley, Wiener, and Zygmund \cite{Dur96b} says that with probability one the paths of the Brownian motion are continuous but not differentiable at every point. Therefore, $V[x_r(u)+B^0_u]$ as a function of $u$ is  not differentiable even if the potential $V$ is. As emphasized by Press \emph{et al.} \cite{Pre92}, higher order quadrature schemes do not automatically translate into better convergence, unless the integrand is well behaved. In our case, there is a limit upon the rate of convergence of the quadrature schemes which is set by the properties of the Brownian motion paths rather than by the properties of the potential, provided that the latter has a continuous first order derivative. In conclusion, one expects that there is an intrinsic limit for the convergence order of the standard DPI methods. Moreover, the Suzuki nonexistence theorem  predicts that none of the classical quadrature formulas for equally spaced abscissas (e.g. Simpson's rules, Bode's rule, etc.) are going to improve upon the asymptotical convergence of the trapezoidal rule. This is strong evidence that the intrinsic limit for the convergence order of the standard DPI methods is 2.

\subsection{Kinetic energy diagonalization for the standard DPI technique}
In practical applications, it is generally difficult to work directly with the formula (\ref{eq:10}) because this involves a correlated Gaussian multidimensional distribution. As shown by Butler and Friedman \cite{But55}, this correlated distribution can be replaced with an independent one by simple algebraic manipulations. Later,  Coalson \cite{Coa86} used a similar technique in order to demonstrate, on an intuitive basis, the relation between the discrete and the Fourier path integral methods. The two approaches mentioned above are technically different and in fact there are  an infinite number of such transformations. As we shall see in this section, they are related by simple orthogonal transformations  and in Section~III we shall propose a new approach, which allows for faster numerical implementations.    

We begin by performing a coordinate transformation so as to diagonalize the positive definite quadratic form associated with the kinetic operator.   More precisely, let us  introduce the transformation of  coordinates $z_n=[x_n -x_r(u_n)]/\sigma$. By using the condition $\sum_i \theta_i =1$, it is straightforward to show that the formula (\ref{eq:10}) becomes
\begin{widetext}
\begin{eqnarray*}
\mathbb{E}\{f[W_{x,x'}^\sigma(u_1), \ldots, W_{x,x'}^\sigma(u_n)]\}=\int_{\mathbb{R}}\ud z_1\ldots \int_{\mathbb{R}}\ud z_n \; f[z_1\sigma+x_r(u_1),\ldots, z_n\sigma+x_r(u_n)]\nonumber \\ \times p_{ \theta_0}(z_1)p_{ \theta_1}(z_2-z_{1})\ldots p_{\theta_{n-1}}(z_n-z_{n-1}) p_{\theta_n}(z_n),
\end{eqnarray*}
\end{widetext}
or, in an even more compact notation, 
\begin{eqnarray*}
\nonumber &&
\mathbb{E}\{f[W_{x,x'}^\sigma(u_1), \ldots, W_{x,x'}^\sigma(u_n)]\}  \\ && =\int_{\mathbb{R}}\ud z_1\ldots \int_{\mathbb{R}}\ud z_n \frac{1}{\sqrt{(2\pi)^n\det (A)^{-1}}} \exp\left({-\frac{1}{2}\bar{z}^{T}A\bar{z}}\right)\nonumber \\ &&\times f[ z_1\sigma+x_r(u_1),\ldots,  z_n\sigma+x_r(u_n)], 
\end{eqnarray*}
where the matrix $A$ is the  $n$-dimensional tridiagonal matrix defined by  $A_{i,i}=1/\theta_i+1/\theta_{i-1}$ for $1\leq i \leq n$ and $A_{i,i+1}=A_{i+1,i}=-1/\theta_i$ for $1\leq i \leq n-1$. 

By construction, the matrix $A$ is symmetric and positive definite [otherwise, the integrability of $\exp\left({-\bar{z}^{T}A\bar{z}}/2\right)$ would be violated] and can be diagonalized by an orthogonal matrix $S$. Defining the new coordinates $\bar{y}=S^T \bar{z}$, and letting $\{\lambda_i;\ 1\leq i \leq n\}$ be the set of the $n$ real and (strictly) positive eigenvalues of $A$, we have  
\begin{eqnarray*}
\nonumber &&
\mathbb{E}\{f[W_{x,x'}^\sigma(u_1), \ldots, W_{x,x'}^\sigma(u_n)]\} =\left[\prod_{i=1}^n \left(\frac{\lambda_i}{2\pi}\right)\right]^{1/2} \\ && \times \int_{\mathbb{R}}\ud y_1\ldots \int_{\mathbb{R}}\ud y_n  \exp\left({-\frac{1}{2}\sum_{i=1}^n \lambda_i y_i^2}\right)\nonumber \\ &&\times f\left[x_r(u_1)+\sigma \sum_{j=1}^n S_{1,j}y_j,\ldots, x_r(u_n)+\sigma \sum_{j=1}^n S_{n,j}y_j \right]. \nonumber
\end{eqnarray*}
Finally, setting $a_i=\lambda_i^{1/2}y_i$, one ends up with
\begin{eqnarray}
\label{eq:12a}\nonumber &&
\mathbb{E}\{f[W_{x,x'}^\sigma(u_1), \ldots, W_{x,x'}^\sigma(u_n)]\} =\int_{\mathbb{R}}\ud a_1\ldots \int_{\mathbb{R}}\ud a_n \\ && \times \left( 2\pi \right)^{-n/2}  \exp\left({-\frac{1}{2}\sum_{i=1}^n  a_i^2}\right)  f\bigg[x_r(u_1)\\ &&+\sigma \sum_{j=1}^n S_{1,j}\frac{a_j}{\lambda_j^{1/2}},\ldots, x_r(u_n)+\sigma \sum_{j=1}^n S_{n,j}\frac{a_j}{\lambda_j^{1/2}} \bigg]. \nonumber
\end{eqnarray}
This formula  is  advantageous for numerical applications because the integration is performed over independent identically distributed Gaussian distributions. Given a quadrature scheme, one diagonalizes the tridiagonal matrix $A$ and tabulates the values of $S_{i,j}$ and $\lambda_i$.
For the case of equally spaced time slices, the eigenvectors and eigenvalues of the matrix $A$ are known analytically:
\[
S_{i,j}=\sqrt{\frac{2}{n+1}}\sin\left(\frac{ij\pi}{n+1}\right),\quad 1\leq i, j \leq n
\] 
and \[\lambda_i=4(n+1)\sin^2\left[\frac{i\pi}{2(n+1)}\right],\quad 1\leq i \leq n, \]respectively.

Similar to the invariance of the Brownian bridge at a change of basis as shown by the Ito-Nisio theorem, the formula (\ref{eq:12a}) is invariant to arbitrary orthogonal transformations of the vectors $\overline{a} =(a_1, \ldots, a_n)$.  Indeed, let $Q$ be an arbitrary $n$-dimensional orthogonal matrix, and consider the coordinate transformation $\overline{a}'=Q^T\,\overline{a}$. Notice that $\sum_i a_i^2 =\sum_i {a'}_i^2$ and define the matrix \begin{equation}
\label{eq:12c}
T_{i,j}=\sum_{k=1}^n \frac{S_{i,k}Q_{k,j}}{\lambda_k^{1/2}}.
\end{equation}
 Then a little algebra shows that 
\begin{eqnarray}
\label{eq:12b}\nonumber &&
\mathbb{E}\{f[W_{x,x'}^\sigma(u_1), \ldots, W_{x,x'}^\sigma(u_n)]\} =\int_{\mathbb{R}}\ud a_1\ldots \int_{\mathbb{R}}\ud a_n \\ && \times \left( 2\pi \right)^{-n/2}  \exp\left({-\frac{1}{2}\sum_{i=1}^n  a_i^2}\right)  f\bigg[x_r(u_1)\\ &&+\sigma \sum_{j=1}^n T_{1,j}{a_j},\ldots, x_r(u_n)+\sigma \sum_{j=1}^n T_{n,j}{a_j} \bigg]. \nonumber
\end{eqnarray}

Because of the additional degrees of freedom, the last formula is more useful in practical applications than the  transformation (\ref{eq:12a}). A good part of the computational time is spent with the evaluation of the current paths. For a monodimensional system, one usually needs a number of operations proportional to $n^2$ in order to compute the vector $T\, \overline{a}$ by matrix multiplication. However, if equally spaced time slices are used, the $n$ elements of the form
$\sum_{j=1}^n S_{i,j}{a_j}/{\lambda_j^{1/2}}$ from Eq.~(\ref{eq:12a})  can be computed by fast sine-Fourier transform in a number of operations proportional to $n\log_2(n)$, provided that $n=2^k-1$ with $k\geq 1$ \cite{Mie01, Pre92a}. Equivalently, one may say that there must be some orthogonal matrix $Q$ such that the associated matrix $T$ defined by the relation (\ref{eq:12c}) is a sparse matrix with at most $k$ nonvanishing elements on any line. Therefore, the evaluation of the elements $T\, \overline{a}$ by direct matrix multiplication requires only $\mathcal{O}(k\cdot n)$ operations.   In this paper, we shall directly find such a matrix $T$ by means of the L\'evy-Ciesielski representation of the Brownian bridge, which is discussed in the next section [see formula (\ref{eq:22b})].

\section{The L\'evy-Ciesielski representation of the Feynman-Ka\c{c} formula} 

As we discussed in the previous section, the transformation (\ref{eq:12a}) was utilized by Coalson in order to establish a connection between the discrete path integral methods and the Wiener-Fourier path integral technique \cite{Coa86}. However, strictly speaking  the Wiener-Fourier sequence of approximations is not equivalent to any discretization scheme. That is, for any $n$, there is no sequence of short-time approximations which by Trotter composition would generate the $n$th order Wiener-Fourier approximation. A more precise statement of this assertion is given at the end of Section III.B. Then, a natural question arises: Is there any random series for which at least a particular subsequence can be thought of as a DPI method? The answer is positive and is furnished by the L\'evy-Ciesielski random series construction of the Brownian bridge. 

In this section, we shall specialize the general theory of the random series representation of the Feynman-Ka\c{c} formula \cite{Pre02} for the particular case of the L\'evy-Ciesielski representation of the Brownian motion. The respective method will be designated by the acronym LCPI.  We shall also derive the three associated methods: the primitive LCPI, the partial averaging LCPI, and the reweighted LCPI. Moreover, with the help of the L\'evy-Ciesielski  series representation, we shall prove the Trotter product rule for the case $n=2^k-1$ and for this subsequence, we shall show that each of the above modifications of the LCPI method can be interpreted as the $n$-order Trotter product of some appropriate short-time approximations. In doing so, we establish a direct connection between the discrete and the random series path integral techniques. As a practical application, we shall obtain a sparse matrix $T$ of the form (\ref{eq:12c}) which requires only $\mathcal{O}(k\cdot n)$ operations to compute the vector $T\,\bar{a}$ by matrix multiplication.  

\subsection{The L\'evy-Ciesielski path integral method.}
\begin{figure}[!tbp] 
   \includegraphics[angle=270,width=8.5cm,clip=t]{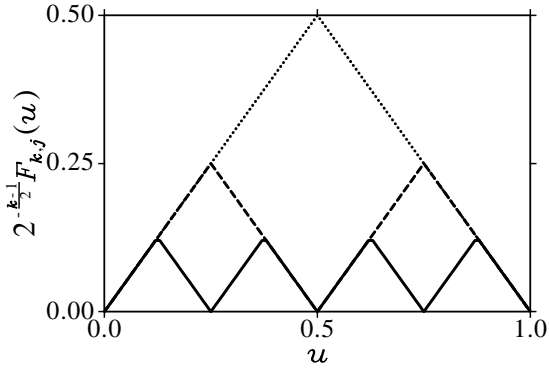} 
 \caption[sqr]
{\label{Fig:2}
A plot of the renormalized Schauder functions for the layers $k=1,2,\,\text{and}\,3$ showing the pyramidal structure.
}
\end{figure}

Some of the arguments we use in the following introduction to the L\'evy-Ciesielski representation of the Brownian bridge  can be found in Ref.~\onlinecite{McK69}. For $k=1,2,\ldots$ and $j=1,2,\ldots,2^{k-1}$, the Haar function $f_{k,j}$  is defined by
\begin{equation}
\label{eq:14}
f_{k,j}(t)=\left\{\begin{array}{cc} 2^{(k-1)/2},& t \in [(l-1)/2^k, l/2^k]\\ - 2^{(k-1)/2},& t \in [l/2^k, (l+1)/2^k]\\ 0, &\text{elsewhere,} \end{array}\right.
\end{equation}
where $l=2j-1$.
Together with $f_0\equiv 1$, these functions make up a complete orthonormal basis in $L^2([0,1])$. Their primitives 
\begin{widetext}
\begin{equation}
\label{eq:15}
F_{k,j}(t)=\left\{\begin{array}{cc} 2^{(k-1)/2}[t-(l-1)/2^k],& t \in [(l-1)/2^k, l/2^k]\\ 2^{(k-1)/2}[(l+1)/2^k-t],& t \in [l/2^k, (l+1)/2^k]\\ 0, &\text{elsewhere} \end{array}\right.
\end{equation}
\end{widetext}
are called the \emph{Schauder functions}. As McKean puts it \cite{McK69}, the Schauder functions are ``little tents,'' which can be obtained one from the other by dilatations and translations. In modern terminology, this has to do with the fact that the original Haar wavelet basis is a multiresolution analysis of $L^2([0,1])$  organized in ``layers'' indexed by $k$ \cite{Mal99}. If we disregard the factor $2^{(k-1)/2}$, the Schauder functions make up a pyramidal structure as shown in Fig.~\ref{Fig:2}.

Let $\{a_{k,j}; k=1,2,\ldots; j=1,2,\ldots,2^{k-1}\}$ be i.i.d. standard normal variables and define $Y_0(u,\bar{a})\equiv 0$ and
\[Y_k(u,\bar{a})=\sum_{j=1}^{2^{k-1}}a_{k,j}F_{k,j}(u).\] Then by Ito-Nisio theorem, 
\begin{equation}
\label{eq:16}
B^0_u(\bar{a})=\sum_{k=1}^{\infty}Y_k(u,\bar{a})
\end{equation} is equal in distribution to a standard Brownian bridge and the convergence of the right hand side random series is uniform almost surely. 

Let us now define the primitive, partial averaged, and reweighted LCPI methods, which are the standard techniques that can be derived from a series representation \cite{Pre02}. They will be denoted in the following discussion by the acronyms Pr-LCPI, PA-LCPI, and RW-LCPI, respectively.  The $n$th order Pr-LCPI term is obtained by approximating the Brownian bridge by the $n$-dimensional process
\[S^n_u(\bar{a})=\sum_{l=1}^{k}Y_l(u,\bar{a})+\sum_{l=1}^{j}a_{k+1,l}F_{k+1,l}(u),\] 
where $k$ and $j$ are the unique numbers such that $n=2^{k}+j-1$, with $k \geq 0$ and $1 \leq j \leq 2^{k}$. However, it appears natural to utilize only the subsequence of the form $n=2^{k}-1$ with $k\geq 0$, corresponding to $k$ complete layers and from now on we shall restrict our attention to this subsequence, for which 
\begin{equation}
\label{eq:17}
S^n_u(\bar{a})=\sum_{l=1}^{k}Y_l(u,\bar{a}).
\end{equation}

Using the notation introduced in Ref.~\onlinecite{Pre02}, we denote the tail of the series (\ref{eq:16}) by
\[\quad \quad B^n_u(\bar{a})=\sum_{l=k+1}^{\infty}Y_l(u,\bar{a}) .\] 
To define the PA-LCPI method, besides the sum (\ref{eq:17}), we need to evaluate
 \[\Gamma_n^2(u)=\sigma^2 \mathbb{E}\, [B^n_u(\bar{a})^2]=\sigma^2\sum_{l=k+1}^\infty \sum_{j=1}^{2^{l-1}} F_{l,j}(u)^2. \]  This quantity must be computed explicitly because it enters the final PA-LCPI formula by means of the ``effective'' potential
\[
\overline{V}_{u,n}(x)=\int_{\mathbb{R}}\frac{1}{\sqrt{2\pi\Gamma_{n}^2(u)}} \exp\left[-\frac{z^2}{2\Gamma_{n}^2(u)}\right]V(x+z) \ud z.
\] 
For more information, the reader is referred to the Section~III of Ref.~\onlinecite{Pre02}.
 For all $l\geq k+1$ and $1\leq j \leq 2^l$, the functions $F_{l,j}$ are zero on the points $u_p=p/2^{k}$ with $p=0,1,\ldots 2^{k}$. Let us define the \emph{support} of the function $F_{l,j}$ as the set $\text{supp}(F_{l,j})=\{u\in[0,1]: F_{l,j}(u)\neq 0\}$.  Moreover, for $1\leq p \leq 2^{k}$, let $I_p=\{(l,j):\; l\geq k+1, \, 1\leq j \leq 2^l, \, \text{supp}(F_{l,j}) \subset [u_{p-1}, u_{p}] \}$ and define 
 \[ W_p(u,\bar{a})=\sum_{(l,j)\in I_p}a_{l,j} F_{l,j}(u). \]
 Then a little thought and the use of the Ito-Nisio theorem show that  $W_p(u,\bar{a})$ is a Brownian bridge  on the interval $[u_{p-1}, u_p]$ of variance $1/2^{k}$. In addition, if $p_1\neq p_2$, then the Brownian bridges $W_{p_1}(u,\bar{a})$ and $W_{p_2}(u,\bar{a})$ are independent because they are functions of the independent Gaussian random variables $\{a_{l,j}\}$ with $(l,j)\in I_{p_1}$ and $(l,j) \in I_{p_2}$, respectively and the sets of indexes  $I_{p_1}$ and $I_{p_2}$ are disjoint. It is convenient to denote by $\mathbb{E}_p$ the conditional expectation over the random variables $a_{l,j}$ with $(l,j) \in I_{p}$. Then, we have 
 \[B^n_u(\bar{a})=\sum_{p=1}^{2^{k}}W_p(u,\bar{a})\] and
 \begin{equation}
 \label{eq:18}\mathbb{E}\,[B^n_u(\bar{a})^2]= \sum_{p=1}^{2^{k}}\mathbb{E}\,[W_p(u,\bar{a})^2]=\sum_{p=1}^{2^{k}}\mathbb{E}_p[W_p(u,\bar{a})^2].\end{equation}
However, one computes
 \begin{equation}
\label{eq:19}
 \gamma^2_{n,1}(u)=\mathbb{E}_1[W_1(u,\bar{a})^2] = \left\{\begin{array}{l l} u(1-2^k u),& 0\leq u < 2^{-k}, \\ 0, & \text{otherwise} \end{array}\right.
 \end{equation}
 and then by translation  \begin{equation}
 \label{eq:20}
\gamma^2_{n,p}(u)\equiv \mathbb{E}_p[W_p(u,\bar{a})^2]= \gamma^2_{n,1}[u-(p-1)/2^{k}].
\end{equation}
Clearly, the functions $\gamma^2_{n,p}(u)$ have \emph{disjoint} support. Finally, Eq.~(\ref{eq:18}) becomes
\begin{equation}
\label{eq:21}
\Gamma^2_n(u)=\sigma^2\sum_{p=1}^{2^{k}}\gamma^2_{n,p}(u),		
\end{equation}
which concludes the definition of the PA-LCPI method.

 The reweighted technique is yet another way of improving the convergence of the primitive method. It has the advantage that it does not require the evaluation of the Gaussian transform of the potential. As discussed in Ref.~\onlinecite{Pre02}, the main idea is to simulate the effect of the partial averaging method by replacing the tail series $B_u^n(\bar{a})$ in the full series expansion  by a collection of random variables $\{R_u^n(b_1,\cdots,b_{n+q})\}_{0 \leq u \leq1}$ defined over an $n+q$ dimensional   probability space ($q$ is a small integer which does not depend upon $n$).  We ask that (i) the variance at the point~$u$ of $R_u^n(b_1,\cdots,b_{n+q})$, denoted by $\Gamma'^2_n(u)$, be as close as possible to $\Gamma_n^2(u)$ and (ii) the variables $S_u^n(a_1,\cdots, a_n)$ and $R^n_u(b_1,\cdots,b_{n+q})$ be independent and their sum have a joint distribution as close to a Brownian 
bridge as possible.  One  candidate for our approach is 
$R_u^n(b_1,\cdots,b_n)=\sum_{p=1}^{n+q} b_p \Omega_p(u),$  where $b_1,\cdots,b_n$ are i.i.d. standard normal random  variables. Condition (ii) above is realized in the Ito-Nisio theorem  by insuring that the collection $\{F_{l,j}(u), \omega_p(u)\}$ with $1 \leq l \leq k$, $1\leq j\leq 2^{l-1}$, and $1\leq p \leq n+q$  
is orthogonal and we shall look for such a collection. Here, $\omega_p(u)$ is the derivative of $\Omega_p(u)$ and it is not required to be normalized. 

As opposed to the Wiener-Fourier series, for the L\'evy-Ciesielski series it is possible to enforce the condition (i) exactly. The analysis done for the partial averaging method showed that we can represent $\Gamma^2_n(u)$ as the sum of $2^k=n+1$ functions of \emph{disjoint} support and which can be obtained one from the other by translation. Intuitively, we  must set $q=1$ and replace the $2^k$ Haar functions making up the $k+1$ layer by
\[\omega_p(u)=\frac{d}{du}\gamma_n[u-(p-1)/2^k].\]
It is easy to notice that the functions $\omega_p(u)$ are orthogonal among themselves because they have disjoint support. Moreover, it is not difficult to see that the Haar functions $f_{l,j}(u)$ are constant on the intervals $[u_{p-1}, u_{p}]$ for all $l \leq k $ 
and therefore, they are orthogonal on the $\omega_p(u)$ functions because
\[\int_{0}^1 \! \omega_p(u) \ud u = \int_{x_{p-1}}^{x_{p}}\!\! \omega_p(u) \ud u = \gamma_{n,1}(1/2^k)-\gamma_{n,1}(0)=0.\]
\begin{figure}[!tbp] 
   \includegraphics[angle=270,width=8.5cm,clip=t]{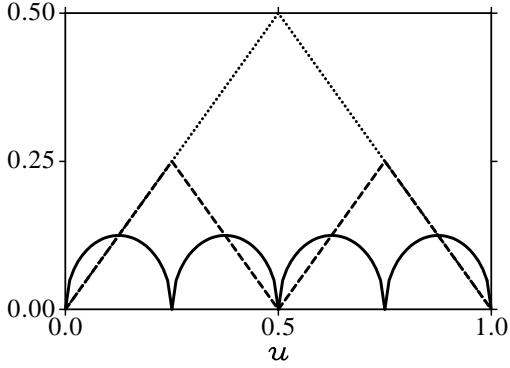} 
 \caption[sqr]
{\label{Fig:3}
A plot of the  functions  used in the reweighted LCPI technique of order $2$. Note that the ``little tents'' of the layer $k=3$ were replaced by ``little domes.''
}
\end{figure}
In consequence, the $n$-order RW-LCPI approximation uses the series
 \begin{equation}
 \label{eq:22}
 S^n_u(\bar{a})=\sum_{l=1}^{k}Y_l(u,\bar{a})+\sum_{p=1}^{2^k}a_{k+1,p}\gamma_{n,p}(u)
 \end{equation}
 for its implementation. 
[It is customary to define the approximation order by the dimensionality of the underlying probability space. We shall not apply this rule in the present paper in order to keep the unity of the exposition.  The squares of the functions $\gamma_{n,p}(u)$ are given by the relation  (\ref{eq:20}).] A look at Fig~\ref{Fig:3} shows that the $n$-order RW-LCPI method is identical to the $2n+1$ order Pr-LCPI, except for the replacement of the last layer of functions $F_{k+1,p}(u)$ with $\gamma_{n,p}(u)$. 

\subsection{Properties of the L\'evy-Ciesielski path integral method.}
As announced in the beginning of the section, the LCPI method for $n=2^k-1$ is virtually a reformulation of the Discrete Path Integral method with appropriate short-time approximations. This is shown by the following result.
\begin{2}
If $n=2^k-1$, then the following relations hold true:
\begin{enumerate}
\item{The Trotter theorem
\begin{eqnarray*}
\rho(x,x';\beta)=\int_{\mathbb{R}}\ud x_1 \ldots \int_{\mathbb{R}}\ud x_n\; \rho\left(x,x_1;\frac{\beta}{n+1}\right)\nonumber \\ \ldots \rho\left(x_n,x';\frac{\beta}{n+1}\right).
\end{eqnarray*}
}
\item{If $Mt$ stands for any of the LCPI methods, then}\end{enumerate}
\begin{eqnarray*}
\rho_n^{Mt}(x,x';\beta)=\int_{\mathbb{R}}\ud x_1 \ldots \int_{\mathbb{R}}\ud x_n\; \rho_0^{Mt}\left(x,x_1;\frac{\beta}{n+1}\right)\nonumber \\ \ldots \rho_0^{Mt}\left(x_n,x';\frac{\beta}{n+1}\right),
\end{eqnarray*}
where the short-time approximations $\rho_0^{Mt}\left(x_n,x';{\beta}\right)$ are defined as follows:
\begin{eqnarray*}
\frac{\rho_0^{Pr}(x,x';\beta)}{\rho_{fp}(x,x';\beta)}=\exp\left\{-\beta \int_0^1 V[x+(x'-x)u] \ud u\right\},
\end{eqnarray*}
\begin{eqnarray*}
\frac{\rho_0^{PA}(x,x';\beta)}{\rho_{fp}(x,x';\beta)}=\exp\left\{-\beta \int_0^1 \overline{V}_{u,0}[x+(x'-x)u] \ud u\right\},
\end{eqnarray*}
and
\begin{eqnarray*}&&
\frac{\rho_0^{RW}(x,x';\beta)}{\rho_{fp}(x,x';\beta)}=\int_{\mathbb{R}}\ud z \left(2\pi\right)^{-1/2}e^{-z^2/2}\\&&\times \exp\left\{-\beta \int_0^1 V[x+(x'-x)u+z\,\Gamma_0(u)] \ud u\right\},
\end{eqnarray*}
respectively.
\end{2}
\emph{Proof.} We only prove the first point of the theorem. It is not difficult to see that the Trotter theorem is in fact a part of  the latter case with $\rho_0^{Mt}(x,x';\beta)=\rho_n^{Mt}(x,x';\beta)=\rho(x,x';\beta)$. As such, the second point follows by arguments similar to the first one and is left to the reader.

Let us remember that  for all $l\geq k+1$ and $1\leq j \leq 2^l$, the functions $F_{l,j}(u)$ are zero on the points $u_p=p/2^{k}$ with $p=0,\ldots 2^{k}$. This means that the joint distribution of the Brownian bridge at these points is \emph{uniquely} determined by the series (\ref{eq:17}). For this proof, it is important to notice that the inverse result is also true: knowledge of the joint distribution of the points $u_p$ with $p=0, \ldots 2^{k}$  uniquely determines the series (\ref{eq:17}) because the latter is \emph{linear} on the intervals $[u_{p-1}, u_p]$. It follows that the variables $B_{u_p}^0$ are independent of the displaced and rescaled Brownian bridges $W_p(u,\bar{a})$.  Using this information together with the joint distribution density for the random variables $x_r(u_p)+\sigma B_{u_p}^0$, which is given by the formula (\ref{eq:9}) as shown in the previous section, one computes  
\begin{widetext}
\begin{eqnarray*}\nonumber
p_{\sigma^2}(x'-x)\mathbb{E}\exp\left\{-\beta\int_{0}^{1}\! \!  V\Big[x_r(u)+\sigma B_u^0 \Big]\ud u\right\} = p_{\sigma^2}(x'-x)\mathbb{E}\exp\left\{-\beta \sum_{p=0}^{n}\int_{u_{p}}^{u_{p+1}}\! \!  V\Big[x_r(u)+\sigma B_u^0 \Big]\ud u\right\}\\ =\nonumber 
\int_{\mathbb{R}} \ud x_1 \cdots \int_{\mathbb{R}} \ud x_{n}\prod_{p=0}^n \left(p_{\frac{\sigma^2}{n+1}}(x_{p+1}-x_p) \mathbb{E}_p \exp\left\{-\beta \int_{u_{p}}^{u_{p+1}}\! \!  V\Big[ x_{p}+\frac{(u-u_p)}{(u_{p+1}-u_p)}(x_{p+1}-x_p) + \sigma W_p(u,\bar{a}) \Big]\ud u\right\}\right)\\=
\int_{\mathbb{R}} \ud x_1 \cdots \int_{\mathbb{R}} \ud x_{n} \prod_{p=0}^{n}\left(p_{\frac{\sigma^2}{n+1}}(x_{p+1}-x_p) \mathbb{E}\exp\left\{-\frac{\beta}{n+1} \int_{0}^{1}\! \!  V\Big[ x_{p}+u(x_{p+1}-x_p) + \frac{\sigma}{\sqrt{n+1}}\, B_u^0(\bar{a}) \Big]\ud u\right\}\right).\qquad
\end{eqnarray*}
\end{widetext}                                
which proves the first point of the theorem. The latter follows by a similar line of thought: for instance, the result for the primitive method is  obtained by setting $W_p(u,\bar{a})=0$ in the previous formula.  $\quad \Box$

From the theoretical point of view, the importance of the Theorem~2 consists of the fact that it establishes a direct connection between the random series and the discrete path integral techniques, even if only for the $n=2^k-1$ subsequence. As such, we notice that the primitive result was employed by Makri and Miller \cite{Mak88,Mak89} and by Mielke and Truhlar \cite{Mie01} as the ZOP-DPI method. The latter authors found that the asymptotic convergence of the method was $O(1/n)$. This result is in good agreement with the present analysis  because  the primitive L\'evy-Ciesielski method cannot exceed the convergence rate of the most rapidly convergent series, the Wiener-Fourier series, which behaves asymptotically as $O(1/n)$ \cite{Pre02}.

The partial averaging result is not new either. The DPI formulation was used by Kole and De~Raedt \cite{Kol01} to treat systems with negative coulombic singularities, for which the non-averaged methods are known to be ill-behaved. However, Kole and De Raedt were not aware of the fact that they were using the partial averaging method in a special setting and regarded their approach as a separate method. It has been shown in a mathematically rigorous way that the partial averaging method is convergent for such potentials at least as far as the pointwise density matrix, the partition function, and related integral expressions are concerned \cite{Pre02b}, for all series representations of the Feynman-Ka\c{c} formula.   Therefore, the Theorem~2 can be used to give a mathematically rigorous proof of the Kole and De~Raedt result, which conversely can be thought of as an argument demonstrating the desirable properties of the partial averaging strategy. 

In a related situation, the partial averaging method as specialized for the Wiener-Fourier series representation was used to treat the polaron problem by Alexandrou, Fleischer, and Rosenfelder \cite{Ale90}.  Later, the DPI formulation of the partial averaging technique was applied by Titantah, Pierleoni, and Ciuchi \cite{Tit01}  for the same polaron problem and regarded once again as a separate technique. We hope we have convinced the reader that given the multitude of series representations that may enter the Feynman-Ka\c{c} formula, there are an infinite number of ways in which the  partial averaging idea can be implemented. The Wiener-Fourier and the L\'evy-Ciesielski series representations as well as the related DPI implementation are only some instances (although perhaps the most important ones). 

In Section II.C, we promised that we would find a quick way to compute the current paths for the standard DPI methods by means of the L\'evy-Ciesielski series representation. For the LCPI formulation, it is straightforward to notice that the computational time necessary to compute the current path at a point $u$ is proportional to $k=\log_2(n+1)$ for the Pr-LCPI and PA-LCPI methods and $1+\log_2(n+1)$ for the RW-LCPI method, respectively. This is so because given a point $u$, the only Schauder function from the layer $l$ that is non-zero at the point $u$ is $F_{l,j}(u)$ with $j=[2^{l-1}u]+1$, where $[x]$ denotes the integral part of $x$. For the RW-LCPI method, we have in addition that the only  function $\gamma_{n,j}(u)$ which is non-zero at the point $u$ is the one with $j=[2^{k}u]+1$. In fact, going back to the proof of Theorem~2, we remember that the joint distribution  of the points $u_p$ with $p=0, \ldots 2^{k}$  uniquely determines the series (\ref{eq:17}) because, in a more mathematical notation, we have  
\begin{equation}
 \label{eq:22a}
 S^n_{u_p}(\bar{a})=B^0_{u_p}(\bar{a})=\sum_{l=1}^{k}Y_l(u_p,\bar{a}),\quad \forall \;1\leq p\leq n.
 \end{equation}
Equation (\ref{eq:22a}) allows us to write the following special form for Eq.~(\ref{eq:12b}):
\begin{widetext}
\begin{eqnarray}
\label{eq:22b}
\mathbb{E}\{f[W_{x,x'}^\sigma(u_1), \ldots, W_{x,x'}^\sigma(u_n)]\}=\mathbb{E}\{f[x_r(u_1)+\sigma  S^n_{u_1}(\bar{a}) , \ldots, x_r(u_n)+\sigma  S^n_{u_n}(\bar{a})]\}\nonumber \qquad \qquad \\ =\int_{\mathbb{R}}\ud a_1\ldots \int_{\mathbb{R}}\ud a_n  \left( 2\pi \right)^{-n/2}  \exp\left({-\frac{1}{2}\sum_{l=1}^k\sum_{i=1}^{2^{l-1}}  a_{l,i}^2}\right)\qquad \qquad \qquad \qquad \qquad \\ \times f\left[x_r(u_1)+\sigma \sum_{l=1}^{k}F_{l,[2^{l-1} u_1]+1}(u_1)a_{l,[2^{l-1} u_1]+1}, \; \ldots\;,x_r(u_n)+\sigma\sum_{l=1}^{k}F_{l,[2^{l-1} u_n]+1}(u_n)a_{l,[2^{l-1} u_n]+1}\right]. \nonumber
\end{eqnarray}
\end{widetext}
This proves that the standard DPI method can be implemented so that the number of operations necessary to compute the current paths is $O(k\cdot n)$. 

As we said at the beginning of the Section III, as opposed to the $n=2^k-1$ subsequence of the L\'evy-Ciesielski representation, no subsequence of  the Wiener-Fourier representation can be rationalized as a DPI method. The precise meaning of this is that if 
\[S^n_{u}(\bar{a})= \sqrt{\frac{2}{\pi^2}}\sum_{k=1}^n a_k \frac{\sin(k\pi u)}{k},\]
then there is no sequence $0=u_0<u_1, \ldots, u_n<u_{n+1}=0$ such that 
\begin{eqnarray}
\label{eq:22c}&&\nonumber
\mathbb{E}\{f[W_{x,x'}^\sigma(u_1), \ldots, W_{x,x'}^\sigma(u_n)]\}\\&=&\mathbb{E}\{f[x_r(u_1)+\sigma  S^n_{u_1}(\bar{a}) , \ldots, x_r(u_n)+\sigma  S^n_{u_n}(\bar{a})]\}\qquad
\end{eqnarray}
for all functions $f(x_1, \ldots, x_n)$. Indeed, remembering $W_{x,x'}^\sigma(u)=x_r(u)+\sigma B^0_u$ and choosing $f(x)=[x-x_r(u_p)]/\sigma^2$ for some interior point $0<u_p<1$, one computes
\[
\mathbb{E}\{f[W_{x,x'}^\sigma(u_1), \ldots, W_{x,x'}^\sigma(u_n)]\}=\mathbb{E}(B^0_{u_p})^2=u_p(1-u_p).
\]
On the other hand, 
\begin{eqnarray*}&&
\mathbb{E}\{f[x_r(u_1)+\sigma  S^n_{u_1}(\bar{a}) , \ldots, x_r(u_n)+\sigma  S^n_{u_n}(\bar{a})]\}\nonumber \\ &=& \mathbb{E}[S^n_{u_p}(\bar{a})]^2=\frac{2}{\pi^2}\sum_{k=1}^n \frac{\sin^2(k \pi u_p)}{k^2}.
\end{eqnarray*}
Clearly, the equality (\ref{eq:22c}) cannot hold because \[u_p(1-u_p)-\frac{2}{\pi^2}\sum_{k=1}^n \frac{\sin^2(k \pi u_p)}{k^2}=\frac{2}{\pi^2}\sum_{k=n+1}^\infty \frac{\sin^2(k \pi u_p)}{k^2}\] does not vanish on the interval $(0,1)$. To prove this, it is enough to notice that the zeros of $\sin^2[(n+1) \pi u_p]$ and $\sin^2[(n+2) \pi u_p]$ are strictly interlaced.

The following theorem, whose proof is left to the reader, provides the necessary and sufficient conditions for an $n$-order term of an arbitrary series to admit a particular $m$-order DPI representation. 
\begin{3}
Let \[\Gamma_n^2(u)=\sigma^2\left[u(1-u)-\sum_{k=1}^n \Lambda_k(u)^2\right]\] and let $0=u_0<u_1<\ldots<u_m<u_{m+1}=1$. Then 
\begin{eqnarray*}
&&\nonumber
\mathbb{E}\{f[W_{x,x'}^\sigma(u_1), \ldots, W_{x,x'}^\sigma(u_m)]\}\\&=&\mathbb{E}\{f[x_r(u_1)+\sigma  S^n_{u_1}(\bar{a}) , \ldots, x_r(u_m)+\sigma  S^n_{u_m}(\bar{a})]\}\qquad
\end{eqnarray*}
for all $f:\mathbb{R}^m \to \mathbb{R}$ if and only if $\Gamma_n^2(u_p)=0$ for all $1\leq p \leq m$.
\end{3}

The fact that the Wiener-Fourier representation cannot be rationalized as a DPI method should not be surprising. Indeed, we presented enough evidence in Ref.~\onlinecite{Pre02} to support the idea that the convergence of the partial averaging and the reweighted Wiener-Fourier path integral methods is $\mathcal{O}(1/n^3)$ for sufficiently smooth potentials. On the other hand, the analysis performed in Section~II suggests that we cannot expect an asymptotic convergence of the DPI methods better than $\mathcal{O}(1/n^2)$. In fact, as we will show in the next subsection, the $n=2^k-1$ subsequence of the PA-LCPI and RW-LCPI methods  can have at most $\mathcal{O}(1/n^2)$ asymptotic convergence. 

\subsection{Convergence of the PA-DPI and of the RW-DPI methods}
In this subsection, we shall study the convergence of the Trotter product formulae having as short-time approximations the partial averaging and the reweighted zero order formulae given in Theorem~2. It is natural to call this methods the PA-DPI and the RW-DPI methods, respectively. In particular, by virtue of Theorem~2, we obtain the asymptotic rates of convergence for the subsequences $n=2^k-1$ of the corresponding LCPI methods. To anticipate, the convergence of the partition function and of the density matrix will be shown to be $\mathcal{O}(1/n^2)$ for both methods if the potential is smooth enough. More precisely, we limit our discussion to the class of potentials introduced in Ref.~\onlinecite{Pre02b}, which are the Kato-class potentials \cite{Chu95} having finite Gaussian transform. In this section, a potential is called smooth if it lies in the local Sobolev space $W^{1,2}_{\text{loc}}(\mathbb{R}^d)$ and if the squares of the potentials and of the first order derivatives have finite Gaussian transform. We remind the reader that the local Sobolev space $W^{1,2}_{\text{loc}}(\mathbb{R}^d)$ is made up of all $L^2_{\text{loc}}(\mathbb{R}^d)$ functions whose first order  distributional derivatives are also $L^2_{\text{loc}}(\mathbb{R}^d)$ functions i.e.,
\[
\int_{D}\left[V(x)^2+\sum_{i=1}^{d}|\partial V(x)/\partial x_i|^2\right] \ud x < \infty 
\]
for all bounded domains $D \subset \mathbb{R}^d$. We warn the reader that the $\mathcal{O}(1/n^2)$ convergence of the density matrix and of the partition function for this class of potentials does not automatically imply similar convergence for the energy estimators, for which additional restrictions upon the class of potentials might be necessary. 

To simplify the notation, we prove the convergence results for the monodimensional case and only state the multidimensional analogues.
Let us start with the asymptotic convergence of the partial averaging method. If we set 
\begin{equation}
\label{eq:23}
U(x,x',\beta;\bar{a})= \int_{0}^{1}\! \!  V\Big[ x_r(u) + \sigma B_u^0(\bar{a}) \Big]\ud u,
\end{equation}
a little algebra shows that
\begin{eqnarray*}
\left|\rho(x,x';\beta)-\rho_0^{PA}(x,x';\beta)\right|=\rho(x,x';\beta)-\rho_0^{PA}(x,x';\beta)\\=\rho_0^{PA}(x,x';\beta) \,\mathbb{E}\left\{e^{-\beta [U(x,x',\beta;\bar{a})-\mathbb{E}\,U(x,x',\beta;\bar{a})]}-1\right\}.
\end{eqnarray*}
The first equality follows from the fact that zero order PA density matrix is always smaller than the true density matrix, according to equation (18) of Ref.~\onlinecite{Pre02}. However, for $\beta$ small, we can expand the exponential in a Taylor series in order to establish the order of the short-time approximation. We have:
\begin{eqnarray}\label{eq:23a}&&\nonumber
\rho(x,x';\beta)-\rho_0^{PA}(x,x';\beta)=\rho_0^{PA}(x,x';\beta)\\&&\times  \bigg\{\frac{\beta^2}{2}\mathbb{E} \, \left[U(x,x',\beta;\bar{a})-\mathbb{E}\,U(x,x',\beta;\bar{a})\right]^2\\&&+ \frac{\beta^3}{3!}\mathbb{E} \, \left[U(x,x',\beta;\bar{a})-\mathbb{E}\,U(x,x',\beta;\bar{a})\right]^3+ \mathcal{O}(\beta^4)\bigg\}. \nonumber 
\end{eqnarray}
Notice that the term of order one in the Taylor expansion cancels, so the asymptotic behavior is dictated by the  variance of the function $U(x,x',\beta;\bar{a})$. However, looking at the expression (\ref{eq:23}), we see that this variance must also decay to zero as $\beta \to 0$, because $\sigma \to 0$. The same is true for the third order moment and a gradient expansion similar to the one performed in Appendix~A for the variance of the function $U(x,x',\beta;\bar{a})$ shows that 
\[
\frac{\beta^3}{3!}\mathbb{E} \,\left[U(x,x',\beta;\bar{a})-\mathbb{E}\,U(x,x',\beta;\bar{a})\right]^3
\]
decays to zero  as fast as $\mathcal{O}(\beta^{4.5})$.

We shall be more careful in establishing a proper bound on the variance of the function $U(x,x',\beta;\bar{a})$ because this will eventually dictate the asymptotic rate of convergence. 
As shown in Appendix~A, we have
\begin{eqnarray}
\label{eq:24}
\beta T_1(x,x';\beta) &\leq& \mathbb{E} \left[U(x,x',\beta;\bar{a})- \mathbb{E}\,U(x,x',\beta;\bar{a})\right]^2 \nonumber \\ &\leq& {\beta} T_2(x,x';\beta),
\end{eqnarray}
where the functions $T_1(x,x';\beta)$ and $T_2(x,x';\beta)$ satisfy the relation
\begin{eqnarray}
\label{eq:25}
T(x,x')= \lim_{\beta \to 0} T_1(x,x';\beta)= \lim_{\beta \to 0} T_2(x,x';\beta)\nonumber \\=\frac{\hbar^2}{m_0}\int_0^1 \! \ud u \!\int_0^1 \! \ud \tau \, \frac{u+\tau-2u\tau-|u-\tau|}{2} \\ \times \nonumber V^{(1)}[x_r(u)]V^{(1)}[x_r(\tau)].
\end{eqnarray}
In particular, the inequalities 
\begin{eqnarray*}
\label{eq:26}&&
\rho_0^{PA}(x,x';\beta)\leq \rho(x,x';\beta)\\&& \leq  \rho_0^{PA}(x,x';\beta)\left[1+\frac{\beta^3}{2}T_2(x,x';\beta)+\mathcal{O}(\beta^{4})\right] 
\end{eqnarray*}
show that the zero order partial averaging formula is of convergence order 2. Therefore, the assertion of Makri and Miller \cite{Mak88} that $\rho_0^{PA}(x,x';\beta)$ is not an order 2 short-time approximation is wrong. 
Also, notice that 
\begin{equation}
\label{eq:27}
T(x,x)=\frac{\hbar^2}{12 m_0} \|\nabla V(x)\|^2,
\end{equation}
because \[\int_0^1 \! \ud u \!\int_0^1 \! \ud \tau \, \frac{u+\tau-2u\tau-|u-\tau|}{2}=\frac{1}{12}.\]

Trotter composing the relation (\ref{eq:24}) $n$ times and noticing that $\mathcal{O}(\beta^4)$ eventually contributes a term decaying as fast as $1/n^3$, it is but a simple task to establish the identity
\begin{widetext}
\begin{eqnarray}
\label{eq:28}&&\frac{\beta^3}{2(n+1)^3}\sum_{j=0}^n \int_{\mathbb{R}}\!\ud x_1\! \int_{\mathbb{R}}\! \ud x_2\, \rho_{j-1}^{PA}\left(x,x_1;\frac{j\beta}{n+1}\right)\nonumber  \rho_0^{PA}\left(x_1,x_2;\frac{\beta}{n+1}\right)\rho_{n-j-1}^{PA}\left(x_2,x';\frac{(n-j)\beta}{n+1}\right) \\&& \times T_1\left(x_1,x_2;\frac{\beta}{n+1}\right) \leq
\rho(x,x';\beta)- \rho_n^{PA}(x,x';\beta)\leq \frac{\beta^3}{2(n+1)^3}\sum_{j=0}^n \int_{\mathbb{R}}\!\ud x_1\! \int_{\mathbb{R}}\! \ud x_2\, \rho_{j-1}^{PA}\left(x,x_1;\frac{j\beta}{n+1}\right)\nonumber \\&&\times \rho_0^{PA}\left(x_1,x_2;\frac{\beta}{n+1}\right)\rho_{n-j-1}^{PA}\left(x_2,x';\frac{(n-j)\beta}{n+1}\right)T_2\left(x_1,x_2;\frac{\beta}{n+1}\right),
\end{eqnarray}
\end{widetext}
 with the understanding that $\rho_{-1}^{PA}(x,x';0)=\delta(x'-x)$. The above inequality is valid to the order of $\mathcal{O}(\beta^4/n^3)$. Now, notice that in the sense of distributions, we have  
 \begin{eqnarray*}
 \lim_{\beta \to 0} \rho_0^{PA}\left(x_1,x_2;\beta\right)T_1\left(x_1,x_2;\beta\right)  = \lim_{\beta \to 0} \rho_0^{PA}\left(x_1,x_2;\beta\right)\\ \times T_2\left(x_1,x_2;\beta\right)=  \delta(x_1-x_2)T(x_1,x_1).
 \end{eqnarray*}
Multiplying it by $2(n+1)^2/\beta^3$ and using the previous observation, the formula (\ref{eq:28}) becomes
\begin{widetext}
\begin{eqnarray*}
&&\frac{1}{(n+1)}\sum_{j=0}^n \int_{\mathbb{R}}\!\ud x_1 \rho_{j-1}^{PA}\left(x,x_1;\frac{j\beta}{n+1}\right)\nonumber  \rho_{n-j-1}^{PA}\left(x_1,x';\frac{(n-j)\beta}{n+1}\right)   T\left(x_1,x_1\right) \\&&\leq
\frac{2(n+1)^2}{\beta^3}\left[\rho(x,x';\beta)- \rho_n^{PA}(x,x';\beta)\right]\leq \frac{1}{(n+1)}\sum_{j=0}^n \int_{\mathbb{R}}\!\ud x_1 \rho_{j-1}^{PA}\left(x,x_1;\frac{j\beta}{n+1}\right)\nonumber \\&&\times \rho_{n-j-1}^{PA}\left(x_1,x';\frac{(n-j)\beta}{n+1}\right)T\left(x_1,x_1\right),
\end{eqnarray*}
\end{widetext}
in the limit that $n$ is large. Again in the same limit, the Riemann sum from the above expression  transforms into an integral on the interval $[0,1]$ and combining everything  we obtain the following theorem
\begin{4}
\begin{eqnarray}
\label{eq:29}
\lim_{n \to \infty} \frac{2(n+1)^2}{\beta^3}\left[\rho(x,x';\beta)- \rho_n^{PA}(x,x';\beta)\right]\nonumber \\=
\frac{\hbar^2}{12 m_0}\int_{0}^1\left\langle x\left|e^{-\theta \beta H}\|\nabla V\|^2e^{-(1-\theta)\beta H}\right|x'\right\rangle \ud \theta 
\end{eqnarray}
\end{4}
It is convenient to write Eq.~(\ref{eq:29}) as
\begin{eqnarray}
\label{eq:30}&&
\rho(x,x';\beta)\approx \rho_n^{PA}(x,x';\beta)+\frac{\hbar^2\beta^3}{24 m_0(n+1)^2}\nonumber \\&&\times
\int_{0}^1\left\langle x\left|e^{-\theta \beta H}\|\nabla V\|^2e^{-(1-\theta)\beta H}\right|x'\right\rangle \ud \theta. 
\end{eqnarray}
The $d$-dimensional version of Theorem~4 can be formally obtained by replacing $\|\nabla V\|^2/m_0$ with
\[
\sum_{i=1}^{d} \frac{1}{m_{0,i}}\left[\frac{\partial}{\partial x_i}V(x_1,\ldots,x_d)\right]^2.
\]

Finally, we turn our attention to the convergence of the RW-DPI method. It was previously proved \cite{Pre02b} that for the class of potentials considered in this section the density matrix and the partition function of any  partial averaging method is convergent to the correct result. However, this might not be true of the primitive and the reweighted methods, as well as of the standard DPI methods. Indeed, it is known that the non-averaged methods suffer from what is called ``classical collapse'' for potentials with negative coulombic singularities \cite{Law69,Kle95, Mus97, Kol01}, for which the partial averaging method is, however, convergent. For such systems it happens that the $n$ order partition functions of the primitive, reweighted, and standard DPI methods are always $+\infty$, yet the true quantum partition function is finite. This situation can be prevented by requiring that the \emph{classical} partition function be finite. For instance, for the case of the primitive random series the Jensen's inequality implies
\begin{eqnarray*}
Z^{Pr}_n(\beta)=\frac{1}{\sqrt{2\pi \sigma^2}} \int_{\mathbb{R}} \ud x \int_{{\Omega}} \ud P(\bar{a}) \exp\Big\{-\beta\int_{0}^{1} V[x\\ +\sigma \sum_{k=1}^{n} a_k \Lambda_k(u)]\ud u\Big\} \leq \frac{1}{\sqrt{2\pi \sigma^2}} \int_{\mathbb{R}} \ud x \int_{{\Omega}} \ud P(\bar{a}) \\ \times \int_{0}^{1} \ud u  \exp\Big\{-\beta V[x+ \sigma \sum_{k=1}^{n} a_k \Lambda_k(u)]\Big\}. 
\end{eqnarray*}
By changing the order of integration, one ends up with 
\begin{equation}
\label{eq:36a}
Z^{Pr}_n(\beta)\leq \frac{1}{\sqrt{2\pi \sigma^2}} \int_{\mathbb{R}} e^{-\beta V(x)}\ud x=Z_{\text{cl}}(\beta) < \infty, 
\end{equation}
which proves our assertion. The inequality (\ref{eq:36a}) holds for the reweighted methods  and the standard DPI methods, too (for the latter techniques one uses the condition $\sum_i w_i=1$ and the discrete analog of the Jensen's inequality). In this paper, the condition $Z_{\text{cl}}(\beta)< \infty$ is assumed to hold any time one deals with the non-averaged methods.

Going back to the asymptotic convergence problem, we may follow the reasoning for the partial averaging method provided that we interpret $\mathbb{E}'$ to mean the average against the Gaussian measure \[\ud \mu(z)= \frac{1}{\sqrt{2 \pi}} e^{-z^2/2}\ud z.\]
By Jensen's inequality one proves the inequality
\begin{widetext} 
\begin{eqnarray*}&&
\frac{\rho_0^{RW}(x,x';\beta)}{\rho_{fp}(x,x';\beta)}=\int_{\mathbb{R}}\ud \mu(z) \exp\left\{-\beta \int_0^1 V[x_r(u)+z\,\Gamma_0(u)] \ud u\right\}\\ && \geq \exp\left\{-\beta \int_0^1  \int_{\mathbb{R}}\ud \mu(z) V[x_r(u)+z\,\Gamma_0(u)] \ud u\right\}= \frac{\rho_0^{PA}(x,x';\beta)}{\rho_{fp}(x,x';\beta)}.
\end{eqnarray*}
\end{widetext}
Therefore, $\rho_n^{RW}(x,x';\beta) \geq \rho_n^{PA}(x,x';\beta)$. Moreover, the following analog of Eq.~(\ref{eq:23a}) holds
\begin{eqnarray*}&&
\rho_0^{RW}(x,x';\beta)-\rho_0^{PA}(x,x';\beta)=\rho_0^{PA}(x,x';\beta)\\&&\times  \bigg\{\frac{\beta^2}{2}\mathbb{E}' \; \left[U'(x,x',\beta;z)-\mathbb{E}'\,U'(x,x',\beta;z)\right]^2\\&&+ \nonumber\frac{\beta^3}{3!}\mathbb{E}' \; \left[U'(x,x',\beta;z)-\mathbb{E}'\,U'(x,x',\beta;z)\right]^3+ \mathcal{O}(\beta^4)\bigg\}, \nonumber 
\end{eqnarray*}
where we now define $U'(x,x',\beta;z)= V[x_r(u)+z\,\Gamma_0(u)]$. 
As discussed in Appendix~A, we have
\begin{eqnarray}
\label{eq:37}
\beta T'_1(x,x';\beta) &\leq& \mathbb{E}' \left[U'(x,x',\beta;z)- \mathbb{E}'\,U'(x,x',\beta;z)\right]^2 \nonumber \\ &\leq& {\beta} T'_2(x,x';\beta),
\end{eqnarray}
where the functions $T'_1(x,x';\beta)$ and $T'_2(x,x';\beta)$ satisfy the relation
\begin{eqnarray}
\label{eq:38}
T'(x,x')= \lim_{\beta \to 0} T'_1(x,x';\beta)= \lim_{\beta \to 0} T'_2(x,x';\beta)\nonumber \\= \frac{\hbar^2}{m_0} \int_0^1 \! \ud u \!\int_0^1 \! \ud \tau \,\sqrt{u(1-u)\tau(1-\tau)}   \\\times V^{(1)}[x_r(u)]V^{(1)}[x_r(\tau)]. \qquad\nonumber
\end{eqnarray}
We also have
\[
T'(x,x)= \frac{\pi^2\hbar^2}{64m_0}\|\nabla V(x)\|^2
\]
because
\[\int_0^1 \! \ud u \!\int_0^1 \! \ud \tau \,\sqrt{u(1-u)\tau(1-\tau)}=\frac{\pi^2}{64}.\]
We leave it for the reader to rework the previous arguments for the partial averaging case and show that for large $n$ we have 
\begin{eqnarray*}&&
\lim_{n\to \infty} \frac{2(n+1)^2}{\beta^3}\left[\rho_n^{RW}(x,x';\beta)- \rho_n^{PA}(x,x';\beta)\right] \\ &&=\frac{\pi^2\hbar^2}{64 m_0}\int_{0}^1\left\langle x\left|e^{-\theta \beta H}\|\nabla V\|^2e^{-(1-\theta)\beta H}\right|x'\right\rangle \ud \theta 
\end{eqnarray*}
Since $\pi^2/64 > 1/12$, the previous result demonstrates that for $n$ large enough  $\rho_n^{RW}(x,x';\beta) \geq \rho(x,x';\beta)$, so that the convergence of the RW-DPI is eventually from above. Combining with Theorem~4, one obtains
\begin{5}
\begin{eqnarray*}
\lim_{n\to \infty} \frac{2(n+1)^2}{\beta^3}\left[\rho(x,x';\beta)- \rho_n^{RW}(x,x';\beta)\right]=
-\frac{\hbar^2}{4 m_0}\\ \times \left(\frac{\pi^2}{16}-\frac{1}{3}\right)\int_{0}^1\left\langle x\left|e^{-\theta \beta H}\|\nabla V\|^2e^{-(1-\theta)\beta H}\right|x'\right\rangle 
\ud \theta.
\end{eqnarray*}
\end{5}

As for the partial averaging case, the statement of Theorem~5 can be written in the short form
\begin{eqnarray}\label{eq:39}\nonumber
\rho(x,x';\beta)\approx\rho_n^{RW}(x,x';\beta)
-\frac{\hbar^2\beta^3}{8 m_0(n+1)^2} \left(\frac{\pi^2}{16}-\frac{1}{3}\right)\\ \times\int_{0}^1\left\langle x\left|e^{-\theta \beta H}\|\nabla V\|^2e^{-(1-\theta)\beta H}\right|x'\right\rangle \ud \theta . \qquad
\end{eqnarray}
From the Theorems (4) and (5) and by using cyclic invariance, one easily proves the following relations:
\begin{6}
\begin{eqnarray*}
\frac{Z(\beta)-Z_{n}^{PA}(\beta)}{Z(\beta)}\approx\frac{\hbar^2\beta^3}{24 m_0(n+1)^2} \frac{ \int_{\mathbb{R}} \rho(x;\beta)\|\nabla V(x)\|^2 \ud x}{\int_{\mathbb{R}} \rho(x;\beta) \ud x}
\end{eqnarray*}
and 
\begin{eqnarray*}
\frac{Z_{n}^{RW}(\beta)-Z(\beta)}{Z(\beta)}\approx\frac{\hbar^2\beta^3}{8 m_0(n+1)^2}\left(\frac{\pi^2}{16}-\frac{1}{3}\right)\nonumber \\\times \frac{ \int_{\mathbb{R}} \rho(x;\beta)\|\nabla V(x)\|^2 \ud x}{\int_{\mathbb{R}} \rho(x;\beta) \ud x}.
\end{eqnarray*}
\end{6}

\emph{Observation~1}
We have ${\pi^2}/{16}-{1}/{3}\approx 0.284 < 1/3$, so one may be tempted to say that the reweighted technique converges at a faster rate than the partial averaging method. However,
as previously mentioned in the text, both the $n$-order LCPI and DPI reweighted techniques actually uses $2n+1$ random variables to parameterize the paths. If the convention of denoting the order of an approximation by the number of variables used to parameterize the paths is obeyed, then the constant ${\pi^2}/{16}-{1}/{3}$ should be increased four times. In this case, we have $4\cdot 0.284 =1.134$ which means that the partial averaging is about $1.134/(1/3)=3.4$ times faster than the reweighted technique. 
 
\emph{Observation~2} The asymptotic relative errors for the partition functions shown in Corollary~1 can be evaluated during the Monte Carlo procedure if so desired. It is a fact established in several occasions \cite{Pre02, Pre02b} that the convergence of the partial averaging density matrix and partition functions for all series representations is monotonically from below. In particular, the PA-DPI subsequence $n=2^k-1$ has the same property since it is identical to the respective subsequence of the PA-LCPI method.  However, it might be possible that the partition function for the reweighted methods are monotonically decreasing from above for the $n=2^k-1$ subsequence. In fact, Golden \cite{Gol65} and Thompson \cite{Tho65} have shown that the partition function for the  subsequence $n=2^k-1$ of the trapezoidal Trotter DPI method is monotonically decreasing and this might be true of the RW-DPI method, too.  

Let us remember that there are potentials, as for instance the potentials with negative coulombic singularities, for which the non-averaged methods do not converge. Conversely, there are smooth and bounded from below potentials, as for instance $V(x)=\exp(x^4)$ for which the non-averaged methods are convergent to the correct result yet the partial averaging method is not convergent because $V(x)=\exp(x^4)$ does not have a finite Gaussian transform. For such potentials, it is expected that the Theorem~5 as well as the second part of the Corollary~1 are still true. 

We shall reinforce the  conclusions of this section by verifying the theorems (4) and (5) for the simple case of the quadratic potential  $V(x)=m_0\omega^2 x^2/2$. Again we use atomic units and set $m_0=1$, $\omega=1$, and $\beta=10$. The evaluation of the $n$-order partial averaging and reweighted elements $\rho_n^{\text{PA}}(0;\beta)$ and $\rho_n^{\text{RW}}(0;\beta)$ is analyzed  in Appendix~B. As discussed in Ref.~\onlinecite{Pre02}, for each method $Mt$ the convergence constant
\[c_{Mt}=\lim_{n \to \infty}\frac{\rho(0;\beta)-\rho_{n}^{Mt}(0;\beta)}{(n+1)^2}\] can be obtained numerically by analyzing the asymptotic slope of the equation
\[
c^{Mt}_n=(4n+2)^2 (n+1/2)\left[{\rho(0;\beta)-\rho_{4n+2}^{Mt}(0;\beta)}\right],
\]
as a function of $n$. More precisely, we have $c_{Mt}=\lim_{n \to \infty} c^{Mt}_{n}-c^{Mt}_{n-1}$.
On the other hand, with the help of the exact density matrix $\rho(x,x';\beta)$ of the quadratic potential \cite{Fey94} and of the relations  (\ref{eq:30}) and (\ref{eq:39}), one computes
\begin{eqnarray*}
c_{PA}=\frac{\beta^3}{24} \int_{0}^1 \ud \theta \int_{\mathbb{R}}\ud x \rho(0,x;\theta\beta)\\ \times \rho[x,0;(1-\theta)\beta]\,x^2 = 0.0713 
\end{eqnarray*}
and $c_{RW}=-[(3\pi^2/16)-1]c_{PA}=-0.0606$, respectively. 
\begin{figure}[!tbp] 
   \includegraphics[angle=270,width=8.5cm,clip=t]{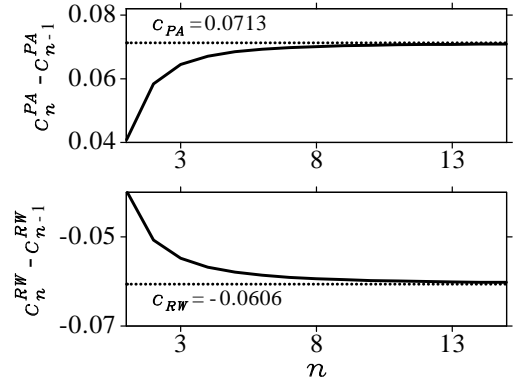} 
 \caption[sqr]
{\label{Fig:4}
The current slopes $c^{Mt}_{n}-c^{Mt}_{n-1}$ (solid lines) are shown  to converge to the values predicted by the theorems (4) and (5) (dotted lines).
}
\end{figure}
The plots in Fig.~\ref{Fig:4} show that indeed, the current slopes $c^{Mt}_{n}-c^{Mt}_{n-1}$ converge to the corresponding values predicted by the theorems~(4) and (5).

To conclude this section, we analyze how smooth  realistic three dimensional  potentials must be to fit the hypothesis of the the theorems (4) and (5). A prototypical example is the tridimensional spherical potential 
\begin{equation}
V(r)=\frac{1}{2}r^2 +\frac{1}{r^\alpha}, \quad  0< \alpha \leq 1,
\end{equation} 
for which  both the partial averaging and the reweighted DPI methods are convergent because the potential $V(r)$ is a Kato-class potential having a finite Gaussian transform and because $Z_{cl}(\beta)< \infty$. The reader may easily verify that $\|\nabla 1/r^\alpha\|^2=\alpha^2/r^{2\alpha+2}$ is locally integrable if and only if  $\alpha < 1/2$. Therefore, if $\alpha< 1/2$,  the theorems (4) and (5) apply and the convergence of both methods is $\mathcal{O}(1/n^2)$. 

On the other hand, if $\alpha \geq 1/2$, the convergence cannot be $\mathcal{O}(1/n^2)$ because the convergence constants are $+\infty$. This can be proved by using the additional information that the density matrix for the Kato-class potentials is continuous and strictly positive. In particular, there is $\epsilon>0$ and $\eta>0$ such that $\rho(x,\beta)\geq \epsilon $ for all $r\leq \eta\leq 1$. Therefore, looking at the bounds for the partition functions given by Corollary~1, we have
\begin{eqnarray*}
\int_{\mathbb{R}^3} \rho(\bar{x};\beta)\|\nabla V(\bar{x})\|^2\ud \bar{x}\geq 4\pi\epsilon \alpha^2  \int_{0}^\eta {r^{-2\alpha}} \ud r \\ \geq 4\pi\epsilon \alpha^2 \int_{0}^1 {r^{-1}} \ud r =+\infty. 
\end{eqnarray*} 
We have treated this problem explicitly in order to show that the nature of the singularities of the potential affects the rate of convergence even if the singularities are oriented ``upward.'' Therefore, in ``pushing'' the Monte Carlo simulation to the limits, the reader may want to actually remove these singularities if they are physically  irrelevant. He/she can do this either by a simple truncation or by approximating the singularity with a better behaved one.  

Other prototypical examples of potentials are those having negative singularities   
\begin{equation}
V(r)=\frac{1}{2}r^2 -\frac{1}{r^\alpha}, \quad  0< \alpha \leq 1.
\end{equation} 
For such potentials, the classical partition function is not finite and the reweighted technique does not properly converge.  However, the partial averaging method is convergent and, if $\alpha <1/2$, the asymptotic convergence is $\mathcal{O}(1/n^2)$. The findings of this section demonstrate that ``smooth enough'' potentials may actually be discontinuous in the three dimensional space.

\section{ Summary and Discussion}
A central theme of  the present paper has been the 
characterization of various path integral approaches and the 
exploration of their interconnections.  The Ka\c{c} interpretation of the 
Feynman approach is a valuable tool for such an analysis. We notice that it is difficult and unnatural to introduce the random series representation by means of the Trotter product rule. Indeed, in order to show that the path integrals
\begin{equation}
\label{eq:fi}
\int_0^1 V[x_r(u)+B_u^0(\bar{a})]\ud u
\end{equation}
are correctly defined, one utilizes the fact that, with probability one, the Brownian paths are continuous.  This property of the Brownian motion is not readily available from the Trotter product rule. However, as we have shown in Section~II, the discrete methods can be directly  derived from the Feynman-Ka\c{c} formula  by simply replacing the  integrals given by Eq.~(\ref{eq:fi}) with appropriate quadrature sums.

We have explored at some length two particular implementations of path 
integral methods: the L\'evy-Ciesielski approach and the associated DPI technique.  We have considered 
primitive, partial averaged and reweighted versions of this 
methods.  As discussed in Section III, the L\'evy-Ciesielski approach 
is of particular importance because its $n=2^k-1$ subsequence can be rationalized both as a 
series and as a discrete method.  This dual character is valuable for 
several reasons.  For example, it provides a convenient and rigorous reformulation of Coalson's findings linking series and discrete path 
integral methods, and, as illustrated by Eq.~(\ref{eq:22b}), suggests a means 
for reducing the numerical overhead associated with path 
construction.  Using the unified framework the L\'evy-Ciesielski 
approach provides, we have shown that the methods introduced by Kole 
and De~Raedt \cite{Kol01} for systems with negative coulombic singularities as 
well as those introduced by Titantah, Pierleoni and Ciuchi \cite{Tit01} for the 
polaron problem are discrete versions of the partial averaging 
approach.  Furthermore, Theorem 2 of Section III suggests that these 
previous methods can be implemented in a more robust manner using the 
L\'evy-Ciesielski series approach.

We have been able to characterize the convergence properties of the 
partial averaging and reweighted DPI approaches and, therefore, of the $n=2^k-1$ subsequence of the corresponding LCPI techniques.  In this respect, Theorems 4 and 5 of Section 
IV provide sharp estimates of the convergence constants for the 
calculation of density matrix elements for both the partial averaged 
and reweighted methods.  To our knowledge, this is the first time 
that such exact convergence constants have been established.  Beyond 
their intrinsic interest, knowledge of these convergence constants 
can be used to devise an improved numerical implementation of the 
Feynman-Ka\c{c} approach.  In particular, the results of Section IV 
indicate that the convergence constants for the reweighted and 
partial averaged methods are related by  the formula $c_{RW}=-[(3\pi^2/16)-1]c_{PA}$ for all pairs of points $(x,x')$ and for all $\beta >0$.  Because the leading 
terms in $1/n^2$  thus cancel, the approach defined by the equation
\begin{equation*}
\rho'_n(x,x';\beta)=\frac{\rho^{\text{RW}}_n(x,x';\beta)+[(3\pi^2/16)-1]\rho^{\text{PA}}_n(x,x';\beta)}{3\pi^2/16}
\end{equation*}
has an asymptotic convergence better than $\mathcal{O}(1/n^2)$, i.e.
\begin{eqnarray*}
\lim_{n\to \infty} \frac{2(n+1)^2}{\beta^3}\left[\rho(x,x';\beta)- \rho'_n(x,x';\beta)\right]=0.
\end{eqnarray*}
In fact, we believe that if the potential $V(x)$ has also a well behaved second derivative, the convergence order of the new method is $\mathcal{O}(1/n^3)$.

Finally, we note that with the help of Theorems 4 and 5, the 
asymptotic behavior of the so-called T-method and H-method energy 
estimators (c.f. Section IV of Ref.~\onlinecite{Pre02}) can be examined.  In 
particular, it should be possible to deduce the convergence constants 
for these estimators from those of the corresponding density matrix 
expressions.  We leave a detailed analysis of such issues for future 
discussion.

\begin{acknowledgments}
The authors acknowledge support from the National Science Foundation through 
awards CHE-0095053 and CHE-0131114. 
\end{acknowledgments}

\appendix
\section{}
It is well known that if $A, B >0$, and $\alpha=C/\sqrt{AB}$ such that $|\alpha|<1$, then the following Mehler's formula \cite{Sze75} holds for all $f$ and $g$ whose squares have finite Gaussian transforms
\begin{widetext}
\begin{eqnarray}
\label{eq:A1}
\overline{[fg]}_{ABC}(x_0,y_0)&=&
\int_{\mathbb{R}}\ud x \!\int_{\mathbb{R}} \ud y \frac{1}{2 \pi} \frac{1}{\sqrt{AB-C^2}} \exp\left(-\frac{1}{2}\frac{x^2 B + y^2 A -2 x y C}{AB-C^2}\right)f(x_0+x)g(y_0+y)\nonumber \\ &=&\int_{\mathbb{R}}\ud x \!\int_{\mathbb{R}} \ud y \frac{1}{2 \pi} \frac{1}{\sqrt{1-\alpha^2}} \exp\left(-\frac{1}{2}\frac{x^2  + y^2  -2 x y \alpha}{1-\alpha^2}\right)f(x_0+x\sqrt{A})g(y_0+y\sqrt{B}) \\&=&
\frac{1}{2\pi} \sum_{k=0}^{\infty} {\alpha^k} \int_{\mathbb{R}}\ud x \!\int_{\mathbb{R}} \ud y \,e^{-(x^2+y^2)/2} \,H_k(x) H_k(y) f(x_0+x\sqrt{A})g(y_0+y\sqrt{B}).\nonumber
\end{eqnarray}
\end{widetext}
In the above, the functions $H_k(x)$ are the normalized Hermite polynomials corresponding to the Gaussian weight 
\[\ud \mu(x)=\frac{1}{\sqrt{2\pi}}e^{-x^2/2}.\] They form a complete orthonormal basis in the Hilbert space $L^2_{\mu}(\mathbb{R})$, which is endowed with the scalar product 
\[\langle \psi| \phi\rangle =\int_{\mathbb{R}} \psi(x)\phi(x)\ud \mu(x).\] Let us notice that  according to our hypothesis, the functions $f(x_0+x\sqrt{A})$ and $g(y_0+x\sqrt{B})$ as functions of $x$ are square integrable against $\ud \mu(x)$ and thus they lie in the Hilbert space $L^2_{\mu}(\mathbb{R})$.

By repeated integration by parts, the formula (\ref{eq:A1}) is shown to equal
\begin{equation}
\label{eq:A2}
\overline{[fg]}_{ABC}(x_0,y_0)=\sum_{k=0}^\infty \frac{C^k}{k!} \overline{f}_A^{(k)}(x_0) \overline{g}_B^{(k)}(y_0),
\end{equation}
where in general $\overline{f}_A^{(k)}(x_0)$ is the $k$-order derivative of 
\[\overline{f}_A(x_0)=\int_{\mathbb{R}} \frac{1}{\sqrt{2\pi A}} e^{-z^2/(2A)}f(x_0+z)\ud z.\]
Let us notice that the series (\ref{eq:A1}) can be extended to the case $\alpha=1$, too. Indeed, the last series in Eq.~(\ref{eq:A1}) for the case $\alpha=1$ is nothing else but the Bessel series  \[\sum_{k=0}^\infty \left\langle H_k|f(x_0+ \cdot \sqrt{A})\right\rangle \left\langle H_k|g(y_0+ \cdot \sqrt{B})\right\rangle,\]
which is convergent to
\begin{eqnarray*}&&
\left\langle f(x_0+ \cdot \sqrt{A})|g(y_0+ \cdot \sqrt{B})\right\rangle \\&& = \int_{\mathbb{R}}f(x_0+ x \sqrt{A})g(y_0+x \sqrt{B})\ud \mu(x). 
\end{eqnarray*}

Next, we proceed to establish the inequality (\ref{eq:24}) from section III.C.
We start with the identity
\begin{eqnarray}
\label{eq:A3}
\mathbb{E} \left[U(x,x',\beta;\bar{a})- \mathbb{E}\,U(x,x',\beta;\bar{a})\right]^2 \nonumber \\ =\mathbb{E}\, U(x,x',\beta;\bar{a})^2- \left[\mathbb{E}\,U(x,x',\beta;\bar{a})\right]^2 .
\end{eqnarray}
Clearly, we have 
\begin{equation}
\label{eq:A4}
\mathbb{E}\,U(x,x',\beta;\bar{a})=\overline{V}_{u,0}[x_r(u)].
\end{equation}
Moreover, 
\begin{eqnarray}
\label{eq:A5}&&
\mathbb{E} \,U(x,x',\beta;\bar{a})^2\nonumber \\&&=\int_0^1\! \ud u \!\int_0^1 \!\ud \tau 
\mathbb{E} \, V[x_r(u)+\sigma B_u^0]V[x_r(\tau)+\sigma B_\tau^0]\qquad
\end{eqnarray}
and the variables $B_u^0$ and $B_\tau^0$ have a joint Gaussian distribution of covariances
\begin{eqnarray*}\mathbb{E}(B_u^0)^2=u(1-u), \ \mathbb{E}(B_\tau^0)^2=\tau(1-\tau)\nonumber \\\text{and} \quad \mathbb{E}(B_u^0B_\tau^0)=\frac{u+\tau-2u\tau-|u-\tau|}{2}.\end{eqnarray*} 
This covariance matrix is independent of any particular representation of the Brownian bridge and therefore can be computed with the help of any basis. For instance, using the Wiener-Fourier basis, the last term of the above formula reads
\begin{equation}
\label{eq:A6}
\mathbb{E}(B_u^0B_\tau^0)=\frac{2}{\pi^2}\sum_{k=1}^\infty \frac{\sin(k\pi u) \sin(k\pi \tau)}{k^2}
\end{equation}
and the sum of the above series can be shown to equal $(u+\tau-2u\tau-|u-\tau|)/{2}$.
It is useful to define the quantities $G_0(u,\tau)=\sigma^2\mathbb{E}(B_u^0B_\tau^0)$ and
\[\Delta^2_0(u,\tau)=\Gamma^2_0(u)\Gamma^2_0(\tau)-G_0(u,\tau)^2.\]
Then,
\begin{eqnarray*}&&
\mathbb{E} \,U(x,x',\beta;\bar{a})^2=\int_0^1\! \ud u \!\int_0^1 \!\ud \tau 
\int_{\mathbb{R}}\ud x \int_{\mathbb{R}} \ud y \frac{1}{2\pi \Delta_0(u,\tau)}\nonumber \\&& \times \exp\left\{-\frac{1}{2}\frac{x^2\Gamma^2_0(\tau)+y^2\Gamma^2_0(u)-2xyG_0(u,\tau)}{\Delta^2_0(u,\tau)}\right\} \\ && \times  V[x_r(u)+x]V[x_r(\tau)+y].\qquad
\end{eqnarray*}

Using the expansion (\ref{eq:A2}), one may write the above integral as the sum of the series
\begin{eqnarray*}
\nonumber
\mathbb{E} \,U(x,x',\beta;\bar{a})^2&=&\sum_{k=0}^{\infty}\frac{1}{k!}\int_0^1\! \ud u \!\int_0^1 \!\ud \tau G_0(u,\tau)^k \\ &\times& \overline{V}_{u,0}^{(k)}[x_r(u)]\overline{V}_{\tau,0}^{(k)}[x_r(\tau)],
\end{eqnarray*}
where $\overline{V}_{u,0}^{(k)}(x)$ is the $k$ order derivative of $\overline{V}_{u,0}(x).$ With the help of Eq.~(\ref{eq:A4}), one recognizes the first term of the above series to be $[\mathbb{E} \,U(x,x',\beta;\bar{a})]^2$, so that  Eq.~(\ref{eq:A3}) becomes
\begin{eqnarray}
\label{eq:A7}&&
\mathbb{E} \left[U(x,x',\beta;\bar{a})- \mathbb{E}\,U(x,x',\beta;\bar{a})\right]^2 \nonumber  =\sum_{k=1}^{\infty}\frac{1}{k!}\\&&\times\int_0^1\! \ud u \!\int_0^1 \!\ud \tau G_0(u,\tau)^k \overline{V}_{u,0}^{(k)}[x_r(u)]\overline{V}_{\tau,0}^{(k)}[x_r(\tau)].\qquad
\end{eqnarray}

Now, we make an important observation: as its eigenfunction expansion (\ref{eq:A6}) shows, $G_0(u,\tau)$ is a positive definite integral kernel $L^2([0,1])\to L^2([0,1])$  and it is not difficult to verify that all $G_0(u,\tau)^k$ are positive definite. Therefore, 
\[\int_0^1\! \ud u \!\int_0^1 \!\ud \tau G_0(u,\tau)^k \overline{V}_{u,0}^{(k)}[x_r(u)]\overline{V}_{\tau,0}^{(k)}[x_r(\tau)]\geq 0\] for all $k\geq 1$. 
Considering only the first term in the series (\ref{eq:A7}), we obtain the inequality
\begin{eqnarray}
\label{eq:A8}
\mathbb{E} \left[U(x,x',\beta;\bar{a})- \mathbb{E}\,U(x,x',\beta;\bar{a})\right]^2 \geq \beta T_1(x,x';\beta)
\end{eqnarray}
where 
\begin{eqnarray*}T_1(x,x';\beta)= \frac{\hbar^2}{m_0}\int_0^1\! \ud u \!\int_0^1 \!\ud \tau \mathbb{E}\,[B^0_u B^0_\tau] \\ \times \overline{V}_{u,0}^{(1)}[x_r(u)]\overline{V}_{\tau,0}^{(1)}[x_r(\tau)].\end{eqnarray*}
It is not difficult to see that as $\beta \to 0$ we have $\Gamma^2_0(u) \to 0$ and so,
\begin{eqnarray}\label{eq:A9}&&
T(x,x')=\lim_{\beta \to 0} T_1(x,x';\beta)= \frac{\hbar^2}{m_0}\nonumber \\ &&\times\int_0^1\! \ud u \!\int_0^1 \!\ud \tau \mathbb{E}\,[B^0_u B^0_\tau]  V^{(1)}[x_r(u)]V^{(1)}[x_r(\tau)]. \qquad
\end{eqnarray} 

To prove the second inequality in Eq.~(\ref{eq:24}), one uses the inequality $1/k! \leq 1/(k-1)!$ and the positivity of the terms of the series (\ref{eq:A7}) to establish the inequality
\begin{widetext}
\begin{eqnarray*}
\mathbb{E} \left[U(x,x',\beta;\bar{a})- \mathbb{E}\,U(x,x',\beta;\bar{a})\right]^2 \nonumber  \leq \sum_{k=0}^{\infty}\frac{1}{k!}\int_0^1\! \ud u \!\int_0^1 \!\ud \tau G_0(u,\tau)^{k+1} \overline{V}_{u,0}^{(k+1)}[x_r(u)]\overline{V}_{\tau,0}^{(k+1)}[x_r(\tau)] \\ = \int_0^1\! \ud u \!\int_0^1 \!\ud \tau G_0(u,\tau) \sum_{k=0}^{\infty}\frac{1}{k!} G_0(u,\tau)^{k} \overline{V}_{u,0}^{(k+1)}[x_r(u)]\overline{V}_{\tau,0}^{(k+1)}[x_r(\tau)] = \int_0^1\! \ud u \!\int_0^1 \!\ud \tau G_0(u,\tau)
\int_{\mathbb{R}}\ud x \int_{\mathbb{R}} \ud y \\ \times \frac{1}{2\pi \Delta_0(u,\tau)} \exp\left\{-\frac{1}{2}\frac{x^2\Gamma^2_0(\tau)+y^2\Gamma^2_0(u)-2xyG_0(u,\tau)}{\Delta^2_0(u,\tau)}\right\}   V^{(1)}[x_r(u)+x]V^{(1)}[x_r(\tau)+y].\qquad
\end{eqnarray*}
\end{widetext}
Therefore, 
\begin{equation}
\label{eq:A10}
\mathbb{E} \left[U(x,x',\beta;\bar{a})- \mathbb{E}\,U(x,x',\beta;\bar{a})\right]^2 \leq \beta T_2(x,x';\beta),
\end{equation}
where 
\begin{eqnarray*}&&
T_2(x,x';\beta)=\frac{\hbar^2}{m_0}\int_0^1\! \ud u \!\int_0^1 \!\ud \tau \mathbb{E}\,[B^0_u B^0_\tau]\int_{\mathbb{R}}\ud x \int_{\mathbb{R}} \ud y \\ &&\times \frac{1}{2\pi \Delta_0(u,\tau)} \exp\left\{-\frac{1}{2}\frac{x^2\Gamma^2_0(\tau)+y^2\Gamma^2_0(u)-2xyG_0(u,\tau)}{\Delta^2_0(u,\tau)}\right\} \\&& \times V^{(1)}[x_r(u)+x]V^{(1)}[x_r(\tau)+y]
\end{eqnarray*}
Again,  as $\beta \to 0$ we have $\Gamma^2_0(u) \to 0$ and
\begin{eqnarray}\label{eq:A11}&&
T(x,x')=\lim_{\beta \to 0} T_2(x,x';\beta)= \frac{\hbar^2}{m_0}\nonumber \\ &&\times\int_0^1\! \ud u \!\int_0^1 \!\ud \tau \mathbb{E}\,[B^0_u B^0_\tau]  V^{(1)}[x_r(u)]V^{(1)}[x_r(\tau)]. \qquad
\end{eqnarray} 

The relations (\ref{eq:A8}),(\ref{eq:A9}),(\ref{eq:A10}), and (\ref{eq:A11}) combined prove the equations (\ref{eq:24}) and (\ref{eq:25}) from Section III.C. The relations (\ref{eq:37}) and (\ref{eq:38}) follow by a similar reasoning and their proof is left to the reader. We only mention that one starts with the fact that the series (\ref{eq:A1}) is well defined and convergent for $\alpha=1$ too, as shown in the beginning of the present appendix. 

\section{}
In this section we discuss the computation of the matrix element $\langle 0| e^{-\beta H}|0\rangle$ for the quadratic potential $V(x)=m_0\omega^2x^2/2$ by means of the standard DPI method and of the partial averaging and the reweighted DPI methods. The density matrix for the quadratic potential is known analytically (see Ref.~\onlinecite{Fey94}) and we do not reproduce it here.  For a standard DPI method specified by the quadrature points $0=u_0< u_1< \ldots< u_n<u_{n+1}=1$, by the increments $\theta_i=u_{i+1}-u_i$, and by the weights $w_0,w_1, \ldots, w_{n+1}$, the formula (\ref{eq:11}) becomes
\begin{eqnarray}
\label{eq:B1}
\rho_{n}^{\text{DPI}}(0;\beta)=\int_{\mathbb{R}}\ud x_1\ldots \int_{\mathbb{R}}\ud x_n \, p_{\sigma^2 \theta_1}(x_1)\nonumber \\ \times p_{\sigma^2 \theta_2}(x_2-x_{1})\ldots p_{\sigma^2 \theta_n}(x_n-x_{n-1}) p_{\sigma^2\theta_n}(x_n)\\ \times \nonumber \exp\left\{-\frac{m_0 \omega^2\beta}{2}\sum_{i=1}^{n} w_i x_i^2\ud u\right\}. 
\end{eqnarray}
Remember that $\sigma^2=\beta \hbar^2/m_0$. If we set $\bar{x}^T=(x_1, x_2, \ldots, x_n)$, the above $n$-dimensional integral can be written in the compact form
\begin{eqnarray}
\label{eq:B2}
\rho_{n}^{\text{DPI}}(0,0;\beta)=\left(\prod_{i=0}^{n}\frac{1}{2\pi\sigma^2\theta_i}\right)^{1/2}\int_{\mathbb{R}^n} e^{-\bar{x}^T A \bar{x}/2}\ud \bar{x}\nonumber \\ = \left(\prod_{i=0}^{n}\frac{1}{2\pi\sigma^2\theta_i}\right)^{1/2}\left[ \det\left(\frac{A}{2\pi}\right)\right]^{-1/2},
\end{eqnarray}
where the matrix $A$ is the tridiagonal matrix defined by 
\[A_{i,i}= \frac{1}{\sigma^2\theta_{i-1}}+\frac{1}{\sigma^2\theta_{i}} + m_0\omega^2 \beta w_i\quad \text{for}\; 1\leq i\leq n\]
and
\[A_{i,i+1}=A_{i+1,i}=-\frac{1}{\sigma^2 \theta_i}\quad \text{for}\; 1\leq i\leq n-1.\]
The values of the quadrature points and the corresponding weights for the trapezoidal rule are well known, while for the Gauss-Legendre quadrature scheme the reader may use the routine given in Ref.~(\onlinecite{Pre92}). 

The zero order partial averaging density matrix for the quadratic potential has the explicit expression
\begin{eqnarray}
\label{eq:B3}&&
\rho_{0}^{\text{PA}}(x,x';\beta)=p_{\sigma^2}(x'-x)\nonumber \\&& \times \exp\left[-\frac{m_0 \omega^2\beta}{6}\left({x^2+{x'}^2+xx'}+{\sigma^2}/{2}\right)\right].
\end{eqnarray}
Using Eq.~(\ref{eq:B3}), the reader may easily deduce that the corresponding $n$-order PA-DPI density matrix is
\begin{eqnarray}
\label{eq:B4}&&
\rho_{n}^{\text{PA}}(0;\beta)= \left(\frac{n+1}{2\pi\sigma^2}\right)^{(n+1)/2} \nonumber \\ &&\times \exp\left[-\frac{\beta^2\hbar^2\omega^2}{12(n+1)}\right]\left[ \det\left(\frac{A}{2\pi}\right)\right]^{-1/2},
\end{eqnarray}
where the tridiagonal matrix $A$ is defined by the relations 
\[A_{i,i}= 2\left[\frac{n+1}{\sigma^2} + \frac{m_0\omega^2 \beta }{3(n+1)}\right]\quad \text{for}\; 1\leq i\leq n\]
and
\[A_{i,i+1}=A_{i+1,i}=-\frac{n+1}{\sigma^2}+\frac{m_0\omega^2 \beta }{6(n+1)}\quad \text{for}\; 1\leq i\leq n-1.\]

Finally, the zero order reweighted density matrix has the form
\begin{eqnarray}
\label{eq:B5}&&
\rho_{0}^{\text{RW}}(x,x';\beta)=p_{\sigma^2}(x'-x)\nonumber  (1+\beta^2 \hbar^2\omega^2/6)^{-1/2}\\&& \times \exp \bigg\{-\frac{m_0 \omega^2\beta}{2}\bigg[\frac{x^2+{x'}^2+xx'}{3}\\&& \nonumber-\frac{\pi^2}{16^2}\frac{\beta^2 \hbar^2\omega^2(x+x')^2}{1+\beta^2 \hbar^2\omega^2/6}\bigg]\bigg\},
\end{eqnarray}
which can be deduced by direct integration. Let us set
\[\eta_n^2=1+\frac{\beta^2 \hbar^2\omega^2}{6(n+1)^2}.\]
Then,
\begin{equation}
\label{eq:B6}
\rho_{n}^{\text{RW}}(0;\beta)=\left(\frac{n+1}{2\pi\sigma^2}\frac{1}{\eta_n^2}\right)^{\frac{n+1}{2}}\left[ \det\left(\frac{A}{2\pi}\right)\right]^{-1/2},
\end{equation}
where the tridiagonal matrix $A$ is defined by the relations 
\[A_{i,i}= 2\left[\frac{n+1}{\sigma^2} + \frac{m_0\omega^2 \beta }{3(n+1)}-\left(\frac{\pi}{16}\right)^2\frac{m_0\hbar^2\beta^3\omega^4}{\eta^2_n(n+1)^3}\right]\]
for $1\leq i\leq n$ and
\[A_{i,i+1}=A_{i+1,i}=-\frac{n+1}{\sigma^2}+\frac{m_0\omega^2 \beta }{6(n+1)}-\left(\frac{\pi}{16}\right)^2\frac{m_0\hbar^2\beta^3\omega^4}{\eta^2_n(n+1)^3}\]
for $ 1\leq i\leq n-1$.

\end{document}